\documentclass[twocolumn,prl,english]{revtex4-2}

\usepackage{amsmath}
\usepackage{amsmath}
\usepackage{amssymb}
\usepackage{graphicx}
\usepackage{subfigure}

\makeatletter

\newcommand{\e}{\mathrm{e}}
\newcommand{\vecr}{\mathbf{r}}
\newcommand{\vecl}{{\bm \ell}}
\newcommand{\vecp}{\mathbf{p}}

\newcommand{\vecv}{\mathbf{v}}

\newcommand{\vecg}{\mathbf{g}}

\newcommand{\Omegav}{\bm \Omega}
\newcommand{\tauv}{\bm \tau}

\newcommand{\D}{\mathrm{d}}
\newcommand{\half}{\frac{1}{2}}

\newcommand{\Drho}{\delta\rho}

\makeatother

\begin{document}

\title{Non reciprocity and odd viscosity in chiral active fluids}

\author{Tomer Markovich$^{1,2}$}

\email{tmarkovich@tauex.tau.ac.il\\}
 
\author{Tom C. Lubensky$^{3}$}
\affiliation{
	$^{1}$School of Mechanical Engineering, Tel Aviv University, Tel Aviv 69978, Israel \\
	$^{2}$Center for Physics and Chemistry of Living Systems, Tel Aviv University, Tel Aviv 69978, Israel \\
	$^{3}$Department of Physics and Astronomy, University of Pennsylvania, Philadelphia, Pennsylvania 19104, USA
}

\date{\today}

\begin{abstract}
	Odd viscosity couples stress to strain rate in a dissipationless way. It has been studied in plasmas under magnetic fields, superfluid ${\rm He}^3$, quantum-Hall fluids, and recently in the context of chiral active matter. In most of these studies, odd terms in the viscosity obey Onsager reciprocal relations. Although this is expected in equilibrium systems, it is not obvious that Onsager relations hold in active materials. By directly coarse graining the kinetic energy and independently using both the Poisson-bracket formalism and a kinetic theory derivation, we find that the appearance of a non-vanishing angular momentum density, which is a hallmark of chiral active materials, necessarily breaks Onsager reciprocal relations. This leads to a non-Hermitian dynamical matrix for the total hydrodynamic momentum and to the appearance of odd viscosity and other non-dissipative contributions to the viscosity. Furthermore, by accounting for both the angular momentum density and interactions that lead to odd viscosity, we find regions in the parameter space in which 3D odd mechanical waves propagate and regions in which they are mechanically unstable. The lines separating these regions are continuous lines of exceptional points, suggesting  a possible non-reciprocal phase transition. 
\end{abstract}

\maketitle

%chiral active matter
Chiral active materials are composed of complex molecules that break both parity and time-reversal symmetry (TRS) at the microscale. This is generally a result of continuous injection of energy and
angular momentum through local torques. Realizations of such fluids are found in a variety of systems across length scales, from nanoscale biomolecular motors~\cite{Goldstein2009,Tabe2003,Petroff2015}, actomyosin networks~\cite{Naganathan2014}, and microscale active colloids~\cite{Snezhko2016,Maggi2015,Irvine2019}, to macroscale-driven chiral grains~\cite{Tsai2005,Carvente2020,Reyes2022}. 
%
% odd viscosity
An important consequence of the breaking of parity and TRS is the possible appearance of {\it odd viscosity}.
Unlike the regular viscosity that dissipates energy, odd viscosity is `reactive'~\cite{odd_review,MarLub2021} and can be obtained using a Hamiltonian theory with no need of adding dissipation~\cite{Abanov2021}. Importantly, Onsager reciprocal relations\footnote{Within linear-irreversible-thermodynamics the thermodynamic fluxes, $J_a=\dot{\psi}_a$, are linear combination of the thermodynamic  forces, $f_a = -\delta F / \delta \psi_a$, such that $J_a = L_{ab}  f_b$. Onsager reciprocal relations states that $L_{ab} = \varepsilon_a \varepsilon_b L_{ba} $ where $\varepsilon_a=\pm 1$ depending on whether $\psi_a$ is even or odd under time-reversal.}~\cite{Onsager,NJP,Mazur,Andelman2021,Mandadapu2020a} predict that such odd viscosity only appears when TRS is broken~\cite{PRX2021,fodor2022,dadhichi2018}, be it due to an external magentic field as was studied in gases~\cite{Kagan-Mak}, plasmas~\cite{Braginskii,LLfluids}, and in the context of quantum-Hall fluids~\cite{Hoyos2014,Read2009,Read2011,Bradlyn2012,Gromov2017}, or due to local torques injected at the particle level~\cite{NJP,MarLub2021,VitelliKubo,Banerjee2017,Rudi2022}.

% total momentum
%When studying chiral active materials there seems to be a need of clarification regarding an ambiguity in defining the stress tensor~\cite{odd_ideal_gas,deG-Prost,Martin}. 
By construction, the stress only enters the dynamics via $\nabla \cdot {\bm \sigma}$, where ${\bm\sigma}$ is the stress tensor. As a result, the dynamics is not modified by the transformation ${\bm \sigma} \to {\bm \sigma} + \nabla \times {\bf B}$~\cite{odd_ideal_gas,deG-Prost,Martin}  (${\bf B}$ is an arbitrary second rank tensor), hence ${\bm\sigma}$ is not unique.
Among the different stress tensors are that associated with center-of-mass (CM) momentum alone and that associated with spin angular momentum (SAM) as well.  We will refer to the latter momentum as the {\it total hydrodynamic momentum} (or hydrodynamic momentum for shortness), with density $\vecg$. This momentum differs from the real total momentum as it does not include motion of rapidly decaying non-hydrodynamic molecular modes~\footnote{Strictly speaking, the hydrodynamic momentum  accounts for the fast modes on average. However, these modes usually result in non-hydrodynamic corrections that are neglected when writing the hydrodynamic equations, see Sec.~II of the SI.}. In passive systems in which the SAM density of rotating molecules (or complex particles) relaxes more rapidly than linear momentum density, there is no difference between the two stress tensors (in the hydrodynamic limit).  However, in chiral active materials in which SAM density is driven, a significantly different behavior may arise.  We argue that the experimentally accessible surface forces are those related to the total hydrodynamic momentum.
%
%Moreover, one can write the stress tensor associated with the center-of-mass (CM) or with the total momentum density, where the latter can always be made symmetric~\cite{Martin}. In passive systems where the spin angular momentum (SAM) density of rotating molecules (or complex particles) relaxes more rapidly than the linear momentum density, there is no difference between the two stress tensors (in the hydrodynamic limit). However, in chiral active systems where the SAM density is driven, a significantly different behavior may arise.
%
%We argue that the experimentally accessible surface forces are those related to the total momentum density $\vecg$, which account for both the center-of-mass (CM) momentum density and the SAM density.

% reciprocity of odd viscosity
In all studies we are aware of, odd terms in the viscosity always obey Onsager reciprocal relations, mainly due to the fact that their origin is usually associated with reciprocal inter-particle collisions, even though these may break parity~\cite{odd_ideal_gas}. Even when the possible existence of such non-reciprocal odd terms is assumed phenomenologically from symmetry arguments~\cite{Abanov2021,VitelliKubo, Mandadapu2020a}, molecular dynamic simulations accounting for the inter-particles interactions~\cite{VitelliKubo} have supported reciprocity. 
%~\footnote{Some studies postulated from symmetry arguments the possibility of non-reciprocal odd terms in the viscosity tensor~\cite{Abanov2021,VitelliKubo}, but any microscopic theory~\cite{MarLub2021}, molecular simulation~\cite{VitelliKubo}, and experiments~\cite{Irvine2019} supported reciprocity.}
%
%
%non reciprocity in active matter
This is quite surprising considering the fact that in active materials, reciprocity need not be obeyed, and non-reciprocity is quite common~\cite{Symmetry_review_2022,Vitelli2021}. Of course microscopic interactions obey Netwton's third law and are reciprocal, but effective interactions mediated by an `active' nonequilibrium medium are not necessarily so, and they are often responsible for the frequent occurrence of non-reciprocity~\cite{Lowen2015,Marchetti2020,Golestanian2020}. The observation that active materials do not obey Onsager reciprocity is not new~\cite{Lau2009,Lowen2015,meredith2020predator,Symmetry_review_2022}, but in this work we find a quite remarkable and simple instance of this -- the mere existence of non-vanishing SAM breaks Onsager reciprocal relations. In passive systems, this has no consequence as angular momentum relaxes  rapidly and does not affect the hydrodynamic equations. However, as we show below, in the presence of active torques this is no longer the case, and reciprocity is generically broken.

% what we did in the PRL
In Ref.~\cite{MarLub2021} we coarse-grained the total hydrodynamic momentum density $\vecg$ and found that it obeys the Belinfante-Rosenfeld relation~\cite{Weinberg1995,Bliokh2021}, which for classical fields means that the total hydrodynamic momentum density equals the CM momentum density plus half of the curl of the internal angular momentum density, $\vecl$~\cite{Martin}. Then, by writing the kinetic energy in the common form $H_k=\int \D\vecr \,\vecg^2/(2\rho)$ we found that odd-viscosity, which obeys  Onsager reciprocal relations, emerges from the Poisson-bracket (PB) formalism~\cite{Lubensky2003,Lubensky2005}. The resultant excitation spectrum exhibits 3D odd mechanical waves, even in non-interacting systems.
%Studying the excitation spectrum of the odd viscosity, novel 3D odd mechanical waves were revealed even in such non-interacting system. 

% this paper   
In what follows, we treat the molecules as rigid bodies that exhibit translational and rotational motion, ignoring finite-frequency non-hydrodynamic molecule modes.  We then coarse-grain the microscopic kinetic energy directly to obtain the well known continuum kinetic energy of a collection of rigid molecules~\cite{Goldstein_book} with additional boundary terms.
%
%In this paper we choose a more direct route and coarse-grain the microscopic kinetic energy directly, \TM{leading to the well known  kinetic energy of a collection of rigid molecules (up to boundary terms)~\cite{Goldstein_book}}. 
Surprisingly we find that these two coarse-graining (CG) methods are not equivalent. 
We believe that the  CG of the Hamiltonian presented here  is correct, and we have supported this finding using a kinetic theory derivation (see Materials and Methods) that does not rely on CG of the Hamiltonian itself.

The CG of this work gives the familiar dynamics of CM and angular momenta, with no odd viscosity~\cite{LubenskyBook,Mazur}. However, when writing the total hydrodynamic momentum dynamics, odd viscosity naturally emerges, and more `odd' terms appear in the viscosity, including one that couples vorticity and pressure (odd pressure~\cite{Abanov2021}). These new viscosity terms exactly cancel terms in the dynamics of the total hydrodynamic momentum and do not allow for propagation of 3D odd mechanical waves. Instead, they require a longitudinal wave to always be accompanied by a transverse wave but not vice versa.
%Indeed, odd viscosity still naturally emerges, but more terms appear in the viscosity including one that couples vorticity and pressure (odd pressure~\cite{Abanov2021}). These new viscosity terms exactly cancel terms in the dynamics of the total momentum and do not allow for propagation of 3D odd mechanical waves. Instead, they require a longitudinal wave to always be accompanied by a transverse wave but not vice versa.
%
Importantly, this means that the dynamics are non-reciprocal, and, indeed, the dynamical matrix for long-wavelength excitations is non-Hermitian. In the presence of an additional odd viscosity that can originate in fluctuations or inter-particle interactions~\cite{VitelliKubo} we find instability lines of exceptional points~\cite{Heiss2012}, which suggests the existence of a non-reciprocal phase transition~\cite{Vitelli2021}.

%%%%%%%%%%%%% Model

\begin{figure}
	\begin{centering}
		\includegraphics[width=0.4\textwidth]{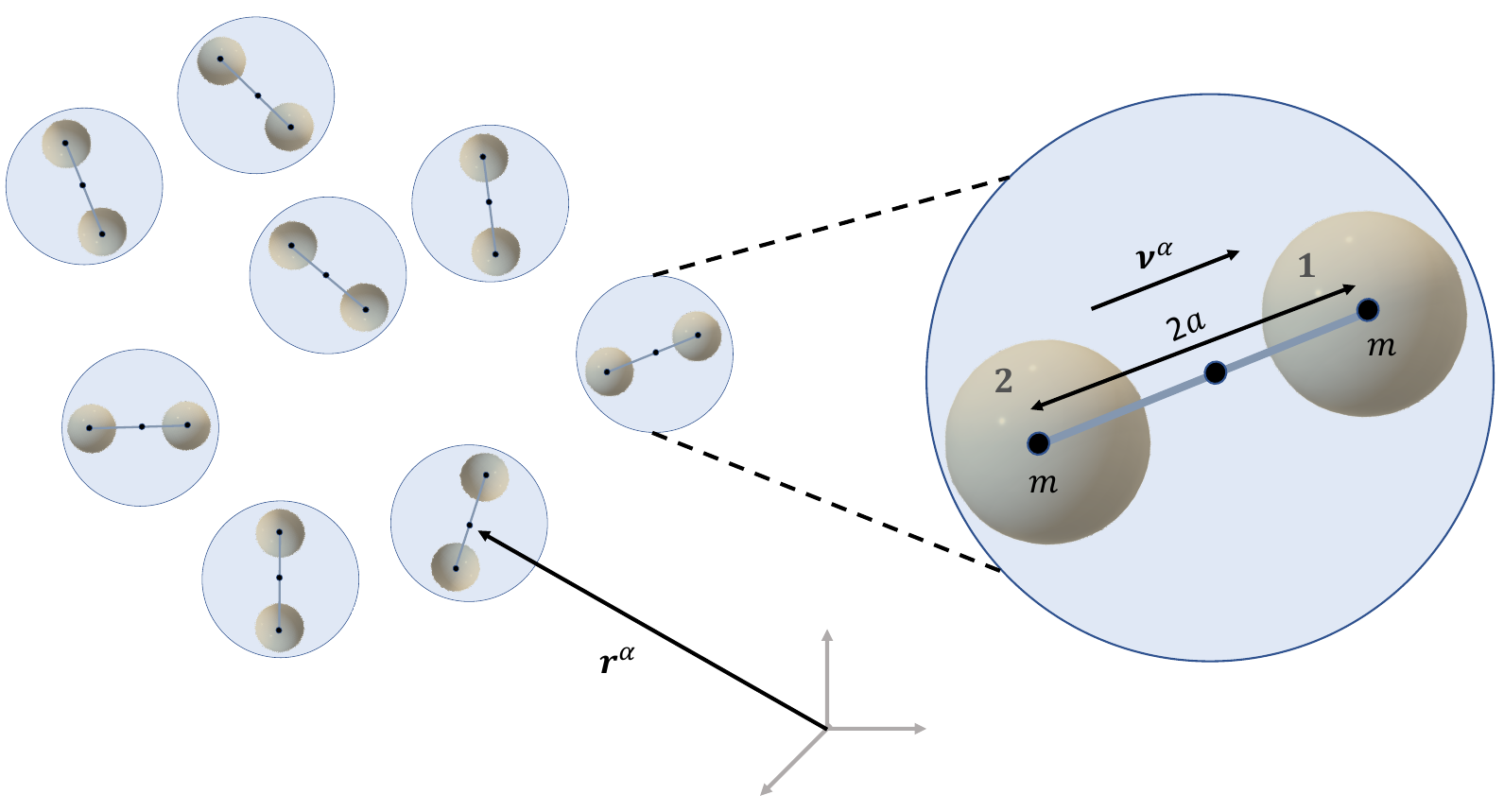}
		\par\end{centering}
	\caption{ Illustration of our model system that is composed of many diatomic molecules. Each of these molecules is composed of two equal point masses, $m$, separated by distance $2a$. The vector ${\bm \nu}$ points from mass `2' to mass `1', and $\vecr^\alpha$ points to the CM of the $\alpha$ molecule. Diatomic molecules are the minimal molecules that has internal angular momentum, and therefore this is the minimal system that will exhibit `kinetic' odd viscosity solely due to molecules spinning, which could be a result of external field or internal {\it active} torques. The existence of angular momentum also breaks Onsager reciprocal relations. 
	}
	\label{fig:diatomic} 
\end{figure}

In what follows we consider a fluid of dumbbells, which are diatomic molecules composed of two point masses, $m$, separated by distance $2a$ as depicted in Fig.~\ref{fig:diatomic}. We define the CM position of  dumbbell $\alpha$ to be $\vecr^\alpha$ while $\bm{\nu}^\alpha$ is a unit vector pointing from mass `2' to mass `1'. 
It is important to note that we only consider a fluid of dumbbells for clearer presentation. The derivation below also applies to general complex rigid molecules (index $\alpha$) that are composed of multiple sub-particles (atoms) with mass $m^{\alpha\mu}$ and momentum $\vecp^{\alpha\mu}$ located at $\vecr^{\alpha\mu}$. We show this in Sec.~III of the  SI. 

\section*{Total hydrodynamic momentum}

We start by introducing the concept of total hydrodynamic momentum, a density field that captures the momentum of all atoms (not only the molecules CM) in the hydrodynamic limit, i.e., in the limit in which all ``fast modes'' are neglected. In practice all finite-frequency modes are discarded and the CG only accounts for the molecules' zero-modes. For simple molecules with no states of self-stress~\cite{Lubensky2015,Mao2018} these zero-modes are the rigid body translations and rotations (see Materials and Methods for more discussion). %in the hydrodynamic limit. 
Hereafter, we will refer to this momentum as the {\it hydrodynamic momentum}.
As we show below, the hydrodynamic momentum flux (or the hydrodynamic momentum stress tensor) is very different from the one associated with the CM. 
%Clearly, the experimentally measurable force in a rheological experiment is the one related to this total momentum stress.

%%%%%%%%%%%% CG total momentum  %%%%%%%%%%%%%%%
To model a fluid of dumbbells,
%When considering a fluid of dumbbells 
we write the momentum of the two point masses in terms of the CM momentum, $\vecp^\alpha$, and ${\bm \nu}^\alpha$ as $\vecp^{\alpha}_{1,2} = \half\vecp^\alpha \pm ma\dot{\bm\nu}^\alpha$, such that the total momentum density for the diatomic fluid, $\hat{g}_i(\vecr) = \sum_{\alpha\mu} p_i^{\alpha\mu} \delta \left(\vecr - \vecr^{\alpha\mu}\right)$, can be split into the CM momentum density, $\hat\vecg^c \equiv \sum_\alpha \vecp^\alpha \delta(\vecr-\vecr^\alpha)$, and a spin-like momentum density, $\hat\vecg = \hat\vecg^c + \hat\vecg^s$, with 
\begin{eqnarray}
	\label{eq:g_s}
	\hat{g}_i^s(\vecr) = \half \nabla \times \hat{\vecl} - \half \nabla\cdot \hat{\bf A} \, ,
\end{eqnarray}
where $\hat{\vecl}(\vecr) = \sum_\alpha \vecl^\alpha \delta(\vecr-\vecr^\alpha)$ and $\hat{{\bf A}}(\vecr) = I\sum_\alpha \dot{\bf Q}^\alpha \delta(\vecr-\vecr^\alpha)$. Here $Q_{ij}^\alpha = \nu_i^\alpha \nu_j^\alpha - \delta_{ij}/d$ is the alignment tensor of molecule $\alpha$ ($d$ is the spatial dimension) and $\hat{Q}_{ij}(\vecr) = \sum_\alpha Q_{ij}^\alpha\delta(\vecr-\vecr^\alpha)$. The angular momentum of each molecule %, $\vecl^\alpha = I {\bm \nu}^\alpha \times \dot{\bm \nu}^\alpha$, 
is related to its angular velocity, $\Omega^\alpha \equiv {\bm \nu}^\alpha \times \dot{\bm \nu}^\alpha$, by $\vecl^\alpha = I {\bm \Omega}^\alpha$. The moment of inertia of each molecule is $I=Ma^2$ with $M=2m$ being the molecule mass. Note that these expressions are written in the long wavelength limit (see Sec.~I of the SI) indicating that we already performed some kind of CG in which the atoms that form the molecule are not considered explicitly. 
Instead, each rigid molecule is described using its CM and angular momenta, which are the generalized momenta associated with the molecule normal zero-frequency modes~\cite{Goldstein_book}.
Therefore, $\hat\vecg = \hat\vecg^c + \hat\vecg^s$ is the {\it total hydrodynamic momentum} and not the real total momentum. 
In principle, one can coarse-grain the toal momentum directly (without taking the hydrodynamic limit at this stage), but then the constraints between the atoms within each molecules will have to be considered explicitly (see Materials and Methods for further discussion). 
%
%This is very useful for the CG of many molecules as now there are no constraints to consider between the normal modes (unlike CG the individual atoms momentum - see Materials and Methods for further discussion).
%%%%%%%%		to the SI
%The difference between our CG (which is the same one used in Ref.\cite{Martin}) and the classical treatment of rigid bodies~\cite{Goldstein_book}, is that our CG takes into account the position of the atoms within the molecule (approximately - in the long wavelength limit) such that a molecule can be part within one CG volume and part in another.
%%%%%%%%%%%
%
%Instead, each molecule is treated as a point-like particle with internal angular momentum (spin). This is very useful as one is usually not interested in the intra-molecule dynamics that relaxes in microscopic times.~\footnote{Consider a diatomic molecule, e.g., ${\rm H}_2$. It has a total of six degrees of freedom: three arising from translation, two from rigid rotation (rotations along the connecting axes are excluded), and one from changes in the separation between the two hydrogen atoms. The latter is a ``fast'' degree of freedom that relaxes rapidly to its equilibrium value, and we ignore it.}

Coarse-graining \eqref{eq:g_s} gives $\vecg^s(\vecr) = \half \nabla\times\vecl - \half\nabla\cdot {\bf A}$. Throughout this work we use a coarse-graining method~\cite{Nakamura2009}  that is described in Materials and Method following ideas of P.~C. Martin.  Assuming the system is in its disordered phase, ${\bf A} = {\bf Q} = 0$, the well-known form of the hydrodynamic momentum density is recovered~\cite{Martin,MarLub2021}.
\begin{eqnarray}
	\label{eq:total_momentum}
	\vecg(\vecr)  = \vecg^c + \frac{1}{2} {\bm \nabla} \times \vecl  \, ,
\end{eqnarray}
where ${\bf g}^c=\rho \vecv^c$ is the CM momentum density, $\rho$ the mass density, $\vecv^c$ the CM velocity, and $\vecl(\vecr)= \rho(\vecr)\tilde{I}\Omega(\vecr)$ is the SAM density,   where $\tilde{I}= I/M = a^2$ is the moment of inertia per unit mass and ${\bm \Omega}$ is the rotation-rate vector. The relation of \eqref{eq:total_momentum} is well known for classical fields~\cite{Martin} and is referred to in the quantum physics litertaure as the Balinfante-Rosenfeld relation in which case the CM momentum is the canonical momentum (derived from Noether theorem) and the angular momentum is the molecular spin~\cite{Bliokh2021}. 

In the classical fields context, Martin {\it et al.}~\cite{Martin} showed that for passive fluids the difference between the total momentum and the CM stress tensors is a microscopic quantity that has no hydrodynamic effects. Our analysis below agrees with this conclusion. However, we find that this is not the case in chiral active fluids, in which case the difference between the stress tensors has hydrodynamic effects and is the source of  odd viscosity. 

\begin{figure}
	\begin{centering}
		\includegraphics[scale=0.3]{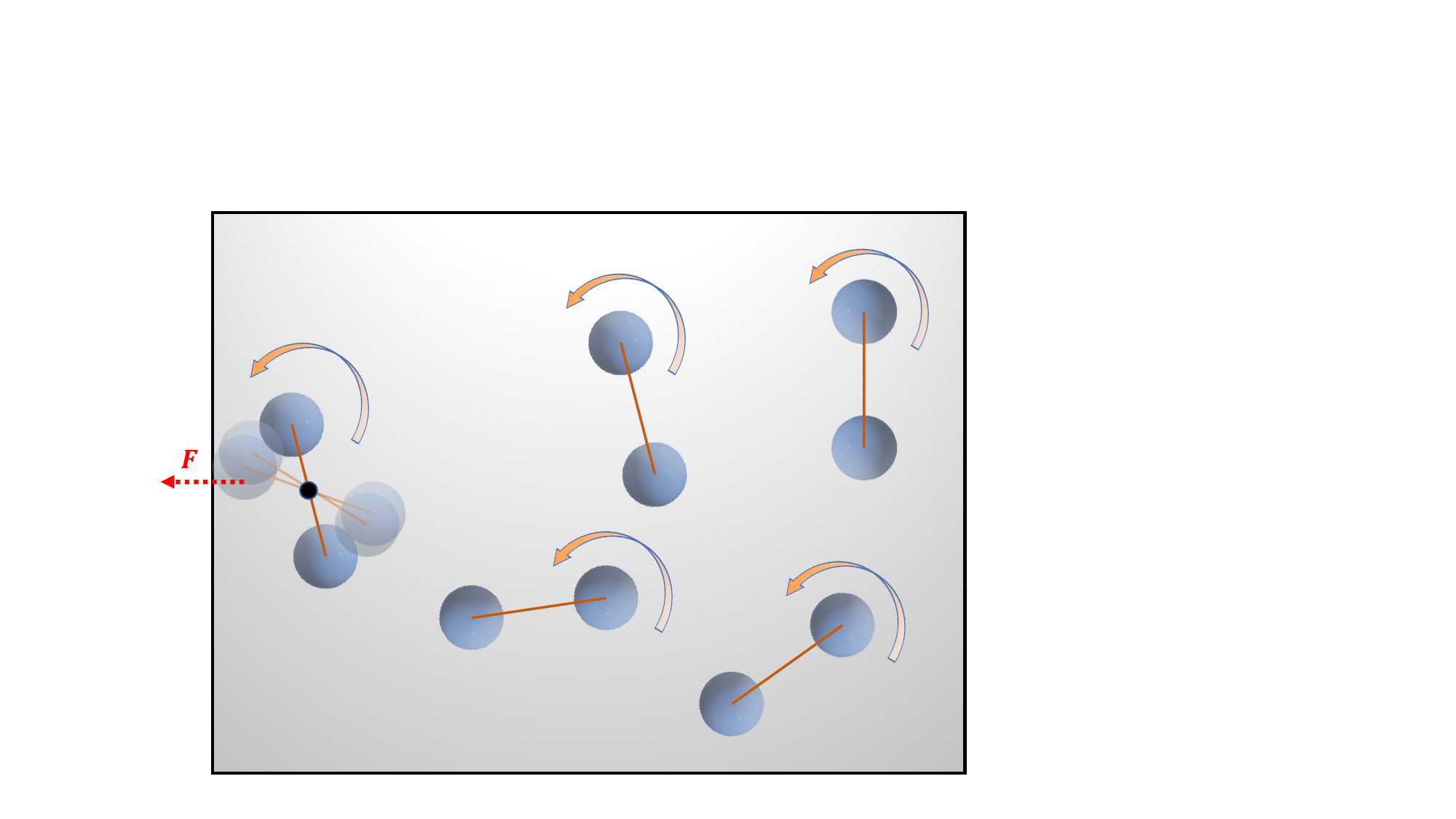}
		\par\end{centering}
	\caption{Cartoon of a fluid of rotating dumbbells. The dumbbells rotate around their CM but their CM do not move. In this specific setting, one of the dumbbells hit the container wall thus exerting a force on it.
		This is perhaps the simplest example of how the CM momentum does not capture all the stress, but the hydrodynamic momentum does.
	}
	\label{fig:total_momentum_stress} 
\end{figure}

Importantly, the macroscopic stress, which is the one accessible experimentally, is the one related to the low-frequency part of the momentum of all atoms.  This is by definition what we call the hydrodynamic momentum stress and not just the CM momentum stress. 
As stated above, for passive fluids this distinction is rather academic. Remarkably, in chiral active fluids this difference is no longer microscopic -- it results in the appearance of odd viscosity and non-reciprocity in the hydrodynamic momentum dynamics. We show this in detail below.
Figure~\ref{fig:total_momentum_stress} illustrates how such difference may arise.

We remark that the real total momentum takes into account the motion of all atoms, which includes all molecules translations, rotations, and internal modes. The CM momentum has a special role as it is also conserved on its own, and as is well known, is sufficient to describe the hydrodynamics of passive fluids, in which all other degrees-of-freedom relax in microscopic times. However, when other, non-hydrodynamic zero-modes are active (or driven) they should be accounted for in the CG process, while finite-frequency modes can usually be neglected (see Sec.~II of the SI for more details).

%\TM{Importantly, the experimentally measurable force in a rheological experiment is the one related to the total momentum stress. Clearly, the macroscopic stress, which is the one accessible experimentally, is the one that is related to the momentum flux of {\it all} atoms. This is by definition the total momentum stress and not just the CM momentum stress. The question is: Why would a macroscopic experiment be sensitive to the microscopic details of the molecules and the difference between the total momentum and CM momentum stress? The remarkable answer is that in passive fluids there is indeed no difference, but in chiral active fluids this difference is the cause of the appearance of odd viscosity and non-reciprocity. We show this in detail below. Figure~\ref{fig:total_momentum_stress} illustrates how such difference may arise.}

%

\section*{Dynamics of the hydrodynamic momentum}

Let us start by restating the well-known dynamics of the CM and angular momenta of isotropic fluids. This can be a simple Newtonian fluid or a complex fluid, but with no order (e.g., no nematic or polar order). 
%
% dynamics of angular momentum
Because we are interested in chiral active fluids, we first consider the dynamics of the SAM density~\cite{NJP,Lubensky2005,MarLub2021,Bartolo2016,Banerjee2017,Mazur}:
\begin{equation}
	\label{e5}
	\dot{\ell}_i(\vecr) + \nabla_j\left(\ell_i v^c_j\right) = - \Gamma \left(\Omega_i - \omega^c_i\right) + \tau_i  \, ,
\end{equation}
where  ${\bm \tau} $ is an external torque density, ${\bm \omega}^c = \half\nabla\times\vecv^c$ is  the rotation vector, and the dissipative term $-\Gamma({\bm \Omega} -{\bm \omega}^c)$ is the one that provides preference for a dissipation-free steady-state in which ${\bm \Omega} = {\bm \omega^c}$, such that the fluid rotates as a rigid body.
The continuity and Navier-Stokes equations are~\cite{Mazur,LubenskyBook}:
\begin{subequations}
	\label{eq:cm_dynamics}
	\begin{align}
		\label{pb1b}
		&\dot\rho + \nabla\cdot\vecg^c = 0 \, , \\
		\label{pb1}
		&\dot{g}_i^c + \nabla_j \left(v_j^c g_i^c\right) =  \nabla_j \Big[ - P \delta_{ij} + \eta^e_{ijkl} \nabla_l v^c_k \nonumber \\ 
		&\qquad\qquad\qquad\,\,\, + \frac{\Gamma}{2}\varepsilon_{ijk} \left(\Omega_k - \omega_k^c\right)\Big]    + f_i  \, ,
	\end{align}
\end{subequations}
where  $\boldsymbol{f}$ is an external force density and $\eta^e_{ijkl} = \lambda\delta_{ij}\delta_{kl} + \eta\left(\delta_{ik}\delta_{jl} + \delta_{il}\delta_{jk}\right)$ is the usual viscosity tensor for an isotropic fluid, with $\lambda$ and $\eta$ being constants. We also include a dissipative antisymmetric stress $\sim\left({\bm \Omega} - {\bm \omega}^c\right)$ with $\Gamma$ being a rotational viscosity~\cite{Mazur,Lubensky2003}, which is required due to the presence of a similar term in \eqref{e5} and conservation of total angular momentum~\cite{NJP}. Here $P = \rho\frac{\delta F}{\delta\rho} - {\cal F}$ is the thermodynamic pressure and $F[\rho]=\int\D\vecr{\cal F}$ is the free energy.

These well-known dynamic equations have been used extensively in literature and clearly show no sign of odd viscosity. Odd viscosity is only revealed upon the realization that the CM momentum density does not account for all of the hydrodynamic momentum.
Taking the time derivative of \eqref{eq:total_momentum} and substituting the dynamics of $\vecg^c$ and $\vecl$ from Eqs.~(\ref{e5})-(\ref{eq:cm_dynamics}) we get the hydrodynamic momentum dynamics (see SI Sec. IV for more details):
%Using the definition of the total momentum, \eqref{eq:total_momentum} and Eqs.~(\ref{eq:cm_dynamics})-(\ref{e5}) to write the dynamics of the total momentum we get:
%
\begin{subequations}
	\label{eq:total_momntum_dynamics}
	\begin{align}
		\label{eq:NSE_total_momentum}
		&\dot{g}_i + \nabla_j(g_i v_g) = \nabla_j \sigma_{ij} + f_i \, ,\\
		\label{eq:total_stress}
		&\sigma_{ij} = -P\delta_{ij} + \left(\eta^o_{ijkl} + \eta^e_{ijkl}\right) \nabla_l v_k + \half\varepsilon_{ijk} \tau_k \, ,\\
		\label{eq:odd viscosity}
		&\eta^o_{ijkl} = -\frac{\ell_n}{4} \left( \gamma^o_{ijkl;n} + \varepsilon_{ilk}\delta_{nj}  + \varepsilon_{jlk}\delta_{ni}  - 2\varepsilon_{lkn}\delta_{ij} \right)  \, ,
	\end{align}
\end{subequations}
with the usual~\cite{Avron1998,MarLub2021,LLfluids,Banerjee2017} odd viscosity tensor 
%
%\TM{discuss symmtery of i->j and total momentum}
%
%
\begin{eqnarray}
	\label{eq:odd_tensor}
	\gamma^o_{ijkl;n} = \varepsilon_{iln}\delta_{jk} + \varepsilon_{ikn}\delta_{jl} + \varepsilon_{jkn}\delta_{il}  + \varepsilon_{jln}\delta_{ik} \, .
\end{eqnarray}
In deriving the above equations we divided the velocity gradient tensor into its symmetric and antisymmetric parts and dropped a non-hydrodynamic term $\sim \nabla \left(\nabla\times\vecl\right)^2$ (see Sec.~IV of the SI). 
%(Hereafter the superscript $b^o$ refers to the `odd' part of $b$.)
%
Importantly, the dissipative term $\sim\left({\bm \Omega} - {\bm \omega}^c\right)$ is canceled in the derivation of \eqref{eq:total_momntum_dynamics} such that one can obtain odd viscosity without the need to introduce dissipation at all. Indeed, the dynamics of the hydrodynamic momentum can be derived using PB directly and the odd terms in the hydrodynamic momentum appear as reactive terms (see Materials and Methods).

Note that $\vecl$ in \eqref{eq:total_momntum_dynamics} is not yet a viscosity, rather it is a dynamical variable that obeys the dynamics of \eqref{e5}  (which also shows that the SAM relaxes in finite time and is therefore not hydrodynamic). 
Moreover, it seems that our simple manipulation, of using the hydrodynamic momentum, leads to the appearance of ``odd'' terms whenever the angular momentum density does not vanish, hence, one may think it appears also in passive fluids. This is of course not the case, and $\vecl$ only contributes to the hydrodynamic equation of the hydrodynamic momentum in the presence of active (or external) torques.

This can be understood as follows. Both TRS and parity are broken by $\vecl$, whereas the presence of $\tauv$, which creates activity and gives both ${\bm\Omega}\simeq{\bm\Omega}^0 = \tauv/\Gamma$ and ${\bm \ell} \simeq {\bm \ell}^0= \underline{\underline{\bf I}}  \cdot \tauv/\Gamma$ steady-state values in the long-wavelength (hydrodynamic) limit (see Sec.~V in the SI), is responsible for the existence of odd terms in \eqref{eq:total_stress}.
(Note that the external torque may include surface friction, e.g., $\tauv = \tilde\tauv - \Gamma^{{\rm ex}} {\bm \Omega}$~\cite{MarLub2021}.)
In the absence of activity, there is no breaking of the continuous rotational symmetry, and  $\Omegav$ relaxes in microscopic times to $\bm\omega$. Then, $\vecl$ in \eqref{eq:total_stress} is $\sim{\bm  \omega} \nabla \vecv \sim \nabla^2$, hence, it is not hydrodynamic and must be omitted from the hydrodynamic equations, Eqs.~(\ref{eq:NSE_total_momentum})-(\ref{eq:odd viscosity}). 
Therefore, odd viscosity only appears in \eqref{eq:total_momntum_dynamics} after relaxation of SAM and in the presence of activity.
For the same reasons, we expect to find odd viscosity in magnetic fluids, such as ferrofluids~\cite{Shliomis1972} and ferronematics~\cite{Brand2003}, in which the magnetization breaks continuous rotational symmetry similarly to the SAM in the chiral active fluid we consider.

Consequently, in the passive case, one cannot distinguish between the total momentum and the CM momentum densities as was also pointed out in Ref.~\cite{Martin}. 
Remarkably, in chiral active materials, this statement is no longer valid, and as we show here (and in Ref.~\cite{MarLub2021}), the hydrodynamic momentum stress is not the same as the CM stress. Then, the stress tensor that is related to the experimentally measured surface forces is the momentum flux of the hydrodynamic momentum, which includes the odd terms.

The hydrodynamic momentum stress of \eqref{eq:total_stress} (after SAM relaxation, $\vecl\to\vecl^0$) contains the odd viscosity, the so-called odd pressure~\cite{Abanov2021} that couples vorticity (or rotations) to pressure, and two more terms that couple vorticity to shears that involve the direction of $\vecl$~\cite{Odd_elasticity}, which do not appear in 2D. Importantly, the existence of these non-dissipative terms does not depend on the free-energy $F$, hence, these terms will appear even in a non-interacting system, e.g., a dilute gas of spinning particles with an average common rotation axis.

In Materials and Methods we have derived the dynamics of Eqs.~(\ref{eq:cm_dynamics})-(\ref{eq:total_momntum_dynamics}) using the PB formalism, which depends on the coarse-grained Hamiltonian. Because we find that the direct CG of the kinetic energy gives a different Hamiltonian from that used in Ref.~\cite{MarLub2021}, and therefore leads to different dynamics, we support our current findings using another, unrelated  derivation of the dynamics. We employed a kinetic theory  that does not rely on the CG of the Hamiltonian to derive the hydrodynamic equations (the kinetic theory derivation is also detailed in Materials and Methods).

%Because we find that the two ways of CG  do not give the same results (compare the result of this section and~\cite{MarLub2021}), we support our current findings using another, unrelated  derivation, of the dynamics using a kinetic theory that is detailed in Materials and Methods.  

%
\begin{figure*}
	\begin{centering}
		\includegraphics[width=1\textwidth]{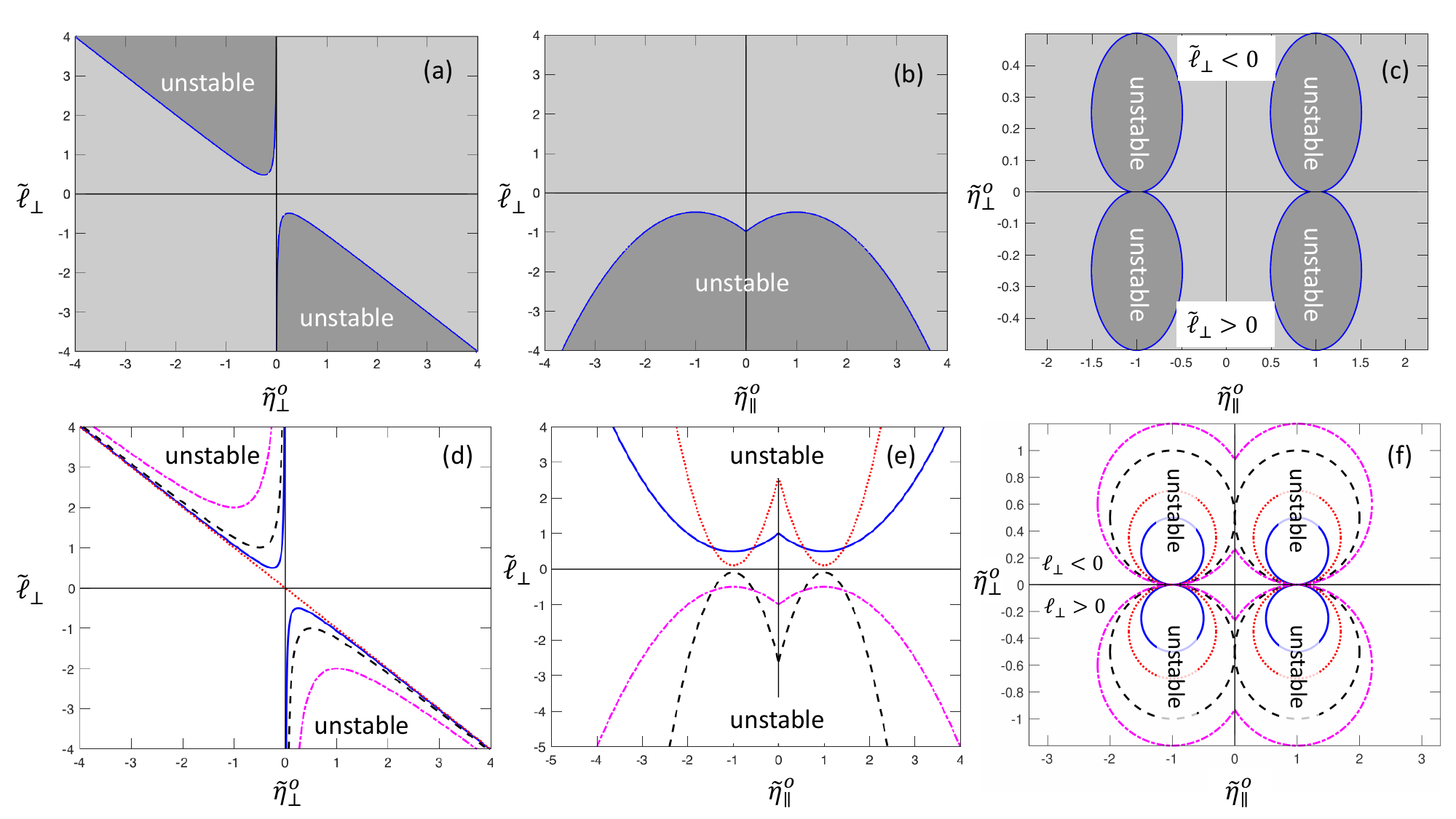}
		\par\end{centering}
	\caption{ 
		Regions of instability in the $(\tilde\ell_\perp,\tilde\eta^o_\perp)$ [for (a) and (d)], $(\tilde\ell_\perp,\tilde\eta^o_\parallel)$  [for (b) and (e)], and $(\tilde\eta^o_\perp,\tilde\eta^o_\parallel)$ [for (c) and (f)] phase-spaces. 
		Instability regions are the dark grey areas in (a), (b), and (c). Lines of exceptional points at which $D=0$ marks the transition, suggesting a non-reciprocal phase transition~\cite{Vitelli2021}.
		A necessary condition for instability is $\tilde\ell_\perp \tilde\eta^o_\perp < 0$, such that there are two instability branches.
		Parameters used are as follows. 
		Top row: (a) $\tilde\eta^o_\parallel = -0.5$, (b) $\tilde\eta^o_\perp = 0.5$ and (c) $\left| \tilde\ell_\perp \right| = 0.5$; 
		Bottom row:
		The lines (solid blue, dotted red, dashed black, magenta dash-dotted) correspond to 
		(d) $\tilde\eta^o_\parallel =  (-0.5,1,2,3)$, (e) $\tilde\eta^o_\perp = (-0.5,-0.1,0.1,0.5)$ , and (f) $\tilde\ell_\perp = (0.5,0.7,1,1.2)$.
		% where the  line colors are blue, red, black, and magenta, respectively. 
	}
	\label{fig:phase} 
\end{figure*}

\section*{Excitation spectrum}

%In this section we calculate the excitation spectrum in the presence of odd viscosity. 
We continue by calculating the excitation spectrum in the presence of odd viscosity. For simplicity we assume that $\tauv$ is constant (e.g., as a result of an external field) 
so that $\vecl\simeq\vecl^0$ is also constant.  We also set {\bf f } = 0 such that the dynamics of the hydrodynamic momentum of \eqref{eq:total_momntum_dynamics} obeys 
%such that $\vecl$ is also constant (see below \eqref{e5}),} and that ${\bf f} = 0$, such that the dynamics for the total momentum obeys
%
\begin{eqnarray}
\label{k7}
\dot{g}_i +  \nabla_j \left(v_j g_i\right) = -\nabla_i \tilde P + \eta \nabla^2 v_i  - \half\varepsilon_{ijn} \ell^0_n \nabla_j \nabla \cdot \vecv  \, ,
\end{eqnarray}
with $\tilde{P} \equiv  P  - (\lambda + \eta) \nabla\cdot\vecv$ being the mechanical pressure (diagonal part of the stress)
\footnote{The odd term of \eqref{k7} seems to be the same as the antisymmetric term in the CM stress $\eta^A$ of Ref.~\cite{odd_review}, but that is not the case. Here, this term comes from a strictly symmetric stress as is written in \eqref{eq:total_momntum_dynamics} for a general $\vecl^0$. Only in the case of constant $\vecl^0$ can one  manipulate the odd terms in \eqref{eq:total_momntum_dynamics} to obtain \eqref{k7}. Importantly, when $\vecl^0$ is constant the antisymmetric stress of \eqref{k7} can be written as a gradient of a third-rank tensor such that total angular momentum is balanced with the active/external torques~\cite{NJP,LLelasticity,Martin}}. (Note that the term $\nabla\times\tauv/2$, which should appear in the right-hand-side of \eqref{k7}, vanishes in the case of constant $\tauv$.)
To find the excitation spectrum we linearize \eqref{k7} and use the Fourier-transform 
$\left[{\bf v},\Drho\right] = \int \frac{\D \bf k}{(2\pi)^3}\left[\tilde{\bf v},\delta\tilde\rho\right] \e^{i\left({\bf k}\cdot\vecr - s t\right)}$, leading to:
\begin{eqnarray}
\label{k8}
\begin{pmatrix}
	s + i\nu k^2  					  & 0   & 2i\ell_r kk_\perp  	&  0\\
	0      & s + i\nu k^2 \,\,  	        &	0		        				& 0 \\
	0        &	0 						    	& s + i\nu_L k^2       		 &  -kc \\
	0 								   	   &	0 					    		&		  -kc 				 	   &  s
\end{pmatrix} 
\begin{pmatrix}
	\tilde{v}_1	\\
	\tilde{v}_2 \\
	\tilde{v}_L \\
	\tilde{h}
\end{pmatrix} = 0 \, . \quad  
\end{eqnarray}
Here $\nu \equiv \eta/\rho_0$, $\nu_L \equiv 2\nu + \lambda/\rho_0$,  $\ell_r \equiv \ell^0/(4\rho_0)$, $\vecl^0 = \ell^0 \hat{z}$, 
and $h = \delta\rho/(c \rho_0)$ where $\rho = \rho_0 + \delta\rho$ and $c$ is the sound speed ($c^2=\partial P/\partial \rho$).
We further define $v_{1,2} \equiv \vecv \cdot \hat{\bf e}_{1,2}$ and $v_L \equiv \vecv \cdot \hat{\bf k}$,
where $\hat{\bf k} = (k_x,k_y,k_z)/k$, $\hat{\bf e}_1 = (-k_y,k_x,0)/k_\perp$, 
and $\hat{\bf e}_2 = (-k_xk_z,-k_yk_z,k_\perp^2)/(kk_\perp)$ are a set of orthonormal vectors with
$k = \sqrt{{\bf k}^2}$, $k_\perp = \sqrt{k_x^2 + k_y^2}$, and $k_\parallel = {\bf k} \cdot \vecl^0 / \ell^0$.

This matrix is not Hermitian, but its eigenvalues can easily be found using the determinant:
\begin{eqnarray}
\label{k9}
\left(s + i\nu k^2\right)^2\left(s^2 - c^2k^2 + is\nu_L k^2\right) = 0 \, .
\end{eqnarray}
The dispersion relation is thus the one found for simple fluids~\cite{LubenskyBook}, where there are two dissipative transverse modes with $s_T=-i\nu k^2$ and two longitudinal decaying waves that obey the usual sound waves relation $s_L = \half k \left[ -i\nu_L k \pm \sqrt{4c^2 - \nu_L^2 k^2} \right]$. The transverse eigenvectors [$\vec{v}=(v_1,v_2,v_L,h)$] are trivial [$\vec{v}_T  = (1, 0, 0, 0) ; (0, 1, 0, 0)$ ] while the longitudinal eigenvectors to lowest order in $g= 2 \ell_r k_{\perp}/c$ and $g_\nu = k\nu_L / c$ are $\vec{v}_L  = (\mp i g, 0,1,\pm1 + ig_\nu/2)$. 

Interestingly, this means that although the presence of non-vanishing SAM  density does not affect the dispersion relation, it creates (a non-reciprocal) coupling, which is evident in the eigenfunctions of \eqref{k8}, between longitudinal and transverse modes such that a longitudinal wave is always accompanied by a transverse wave (but not vice-versa). 

The result of this work stands in contrast to Ref.~\cite{MarLub2021} that predicted the presence of SAM density itself is sufficient to produce a new type of transverse `odd' mechanical waves. It is, nevertheless, expected that spin-spin interactions (that were neglected in this work so far) will result in odd viscosity (that obeys Onsager reciprocal relations)~\cite{VitelliKubo} such that \eqref{eq:odd viscosity} will have an additional term $(\eta^o_n/4)\gamma^o_{ijkl;n}$. Because this `interaction odd viscosity' is a result of spin-spin interaction~\cite{VitelliKubo}, which together with $\vecl$ defines the broken symmetry direction of the system, we expect that ${\bm \eta}^o \parallel \vecl$~\footnote{One could imagine effects of angular momentum biaxiality, but we are not interested in those in this work.}. Then, assuming that $\tauv$, $\vecl$, and ${\bm \eta}^o$ are constants (and ${\bf f}=0$), the linearized equations in the Fourier space, \eqref{k8}, become
\begin{eqnarray}
\label{k10}
\begin{pmatrix}
	s + i\nu k^2  					  \!\!&  -i \nu^okk_\parallel   \!& 2i kk_\perp\!  (\ell_r + \nu^o)	\!\!\!&  0\\
	i \nu^okk_\parallel       \!\!& s + i\nu k^2   	        \!&	0		        			\!\!\!& 0 \\
	-2i \nu^okk_\perp         \!\!&	0 						    	\!& s + i\nu_L k^2       		 \!\!\!&  -kc \\
	0 								   	   \!\!&	0 					    		\!&		  -kc 				 	   \!\!\!&  s
\end{pmatrix} \!\!\!
\begin{pmatrix}
	\tilde{v}_1	\\
	\tilde{v}_2 \\
	\tilde{v}_L \\
	\tilde{h}
\end{pmatrix} \!=\! 0  \, , \nonumber \\  
\end{eqnarray}
where $\nu^o = \left({\bm\eta}^o \cdot \hat{z}\right)/(4\rho_0)$. This matrix is again non-Hermitian, and when dissipation is negligible, $\nu=\nu_L=0$, the dimensionless eigenvalue equation is
\begin{eqnarray}
\label{k11}
x^4 - x^2\left[1 +(\tilde\eta_\parallel^o)^2 + 4\tilde\eta^o_\perp\left( \tilde\eta^o_\perp + \tilde\ell_\perp\right) \right] + (\tilde\eta_\parallel^o)^2  = 0 \, , \quad
\end{eqnarray}
where we define $x=s/(kc)$, $\tilde\eta_\parallel^o = \nu^o k_\parallel/c$, $\tilde\eta_\perp^o = \nu^o k_\perp/c$, and $\tilde\ell_\perp = \ell_r k_\perp/c$. The solutions for this equation can be be written as $x^2 = \left[C \pm \sqrt{D}\right]/2$ where $C = 1 +(\tilde\eta_\parallel^o)^2 + 4\tilde\eta^o_\perp\left( \tilde\eta^o_\perp + \tilde\ell_\perp\right)$ and $D=C^2 \!\!\!- 4(\tilde\eta^o_\parallel)^2$.
%%
%\begin{eqnarray}
%\label{k12}
%\nonumber& &x^2 = \half\left[\chi \pm \sqrt{D}\right] \, ,\\
%\nonumber& &\chi = 1 +(\eta_\parallel^o)^2 + 4\eta^o_\perp\left( \eta^o_\perp + \ell_\perp\right) \, ,\\
%& &D=\chi^2 \!\!\!- 4(\eta^o_\parallel)^2 \, .
%\end{eqnarray}
%%
If the discriminant $D\geq 0$ then $x^2$ is real. If also $C > 0$ then $x^2>0$ and there are four modes of non-decaying waves. When $D\geq 0$ but $C<0$, we have $x^2<0$ and $x$ is pure imaginary, such that there are two modes of non-decaying waves and two unstable modes. %(if $D=0$ there are only two degenerate modes). 
When the discriminant is negative $x^2$ becomes a complex number and $x = \Re\{x\} + i\Im\{x\}$ with $\Re\{x\} = \pm\half\left[  \left(\sqrt{|D|+C^2}+C\right)\right]^{1/2}$ and $\Im\{x\} = \pm\half \left[\sqrt{|D|+C^2}-C\right]^{1/2}$, 
which means that there are two unstable modes with positive imaginary part. Therefore if $D<0$ or $C<0$ the system is linearly unstable, hence the onset of instability is given by $D=0$. %or $\chi < 2\left|\eta^o_\parallel\right|$. 
A special case is when $\tilde\eta^o_\parallel = 0$, for which there are only two modes of non-decaying waves (the regular sound waves) and two zero-modes.

The various instability regions are plotted in Fig.~\ref{fig:phase} in the $(\tilde\ell_\perp,\tilde\eta^o_\perp)$ [for (a) and (d)], $(\tilde\ell_\perp,\tilde\eta^o_\parallel)$  [for (b) and (e)], and $(\tilde\eta^o_\perp,\tilde\eta^o_\parallel)$ [for (c) and (f)] phase-spaces. %From \eqref{k12} it is clear that 
A necessary condition for instability is $\tilde\ell_\perp \tilde\eta^o_\perp < 0$, such that there are two instability branches as shown in Fig.~\ref{fig:phase}. 
%In other regions there are four propagating modes, where on the lines that separate instability regions there are two degenerate modes.
In the stable regions there are four propagating modes.  When $\tilde\ell_\perp = 0$, we recover the results of~\cite{MarLub2021} in which odd waves always propagate and reciprocity is restored.

Remarkably, for any non-vanishing $\tilde\ell_\perp$, the lines of $D=0$ that separate stable and unstable regions are lines of exceptional points~\cite{Heiss2012}. Along these lines, instead of having four propagating modes (two pairs of $\pm$ eigenvalues), each pair of eigenvectors (and eigenvalues) collapse into one, leaving only two propagating modes. Recently, it has been suggested that such exceptional lines mark a new type of phase transition denoted as {\it non-reciprocal phase transition}~\cite{Vitelli2021}. However, our system does not precisely fall within the class of theories examined in~\cite{Vitelli2021} as here there is no spontaneous symmetry breaking and the exceptional points have non-zero eigenvalues.  
It is worth noting that in these exceptional points there are two other modes, but they are not of the form of regular normal modes (see Sec.~VII of the SI for details), instead they have the following form: $\left({\bf W} + (t+c){\bf V} \right)\e^{i\left({\bf k} \cdot \vecr - st\right)}$, where ${\bf V}$ is an eigenvector of the dynamical matrix and ${\bf W}$ is a generalized eigenvector\footnote{A generalized rank-2 eigenvector ${\bf W}$ of a matrix $\underline{\underline{{\bf M}}}$ obeys $\left[\underline{\underline{{\bf M}}} -\lambda \underline{\underline{\bf I}}\right] \cdot  {\bf W} = {\bf V}$, where ${\bf V}$ is a regular eigenvector. Note that $\tilde{\bf W} = {\bf W} + c{\bf V}$ is also a generalized eigenvector and it is orthogonal to ${\bf V}$ if $c=-{\bf W}\cdot{\bf V}/{\bf V}^2$.}
%$\tilde{\bf W}\cdot{\bf V}=0$.}, and $c$ is a constant.  
of the dynamical matrix in Eq.~\ref{k10}. These solutions grow slowly (algebraically) with time and seem unstable, but may be stabilized in the presence of viscosity such that the solution grows at short times and decays at long times, making the system stable along the exceptional lines.
%(algebraically) with time, but may be stabilized in the presence of viscosity, such that the system may be stable along the exceptional lines.
%
% at short times but decay exponentially at long times if $\Re\{s\} >0$.  They may, however be stabilized \TM{even when $\Re\{s\} < 0$} in the presence of viscosity, such that the system may be stable along the exceptional lines.

\section*{Conclusions}

%Odd viscosity was investigated thoroughly in the past few decades in various contexts, from plasmas in magnetic fields, superfluid ${\rm He}^3$, and quantum Hall fluids, to more recent advances in the  %context of chiral active matter. In all of these examples odd viscosity obeyed Onsager reciprocal relations.  

% non reciprocal odd viscosity 
%
The study of chiral active matter has increased interest in odd viscosity dramatically in recent years, from pure theoretical work~\cite{MarLub2021,Abanov2018,Abanov2021,Banerjee2017} to molecular dynamic simulations~\cite{VitelliKubo} and experimental work~\cite{Irvine2019}. %In most of these works odd terms in the viscosity obeyed Onsager reciprocal relations. 
So far, and quite surprisingly, there is no evidence that odd terms in the viscosity breaks Onsager relations.
%
%Some of these works postulate the possibility of odd terms in the viscosity breaking Onsager relations~\cite{Vitelli2021,Abanov2021}, but molecular-dynamics simulations~\cite{VitelliKubo} and experiments~\cite{Irvine2019} show no evidence of that, and there is no a microscopic theory that predicts it. Indeed, in equilibrium systems the reciprocity of Onsager relations must be obeyed, however, in active materials that operates out of equilibrium there is no reason to assume reciprocity. In fact, it is more common to observe lack of reciprocity in such systems. Within this perspective it is actually surprising that odd viscosity appears to obey Onsager relations in chiral active materials.
%
% total momentum
%
In our recent work~\cite{MarLub2021} we  proposed a microscopic model for odd viscosity in active matter, which reproduced an odd viscosity that obeys Onsager relations in the hydrodynamic momentum stress tensor. For that purpose we have used the coarse-grained hydrodynamic momentum density, which accounts for both the center-of-mass linear momentum density and the SAM density. 
%Importantly, when measuring surface forces in a fluid with spinning particles, these must be determined using the stress tensor of the total momentum density rather than the one related to the CM momentum density. 

% introducing \ell breaks Onsager reciprocal relations
% ORR are obeyed in CM but not in total momentum
%
In this paper we directly coarse-grain the kinetic energy which results in somewhat different results. The reactive part of the CM stress tensor does not contain any trace of the SAM density and therefore has no odd viscosity. It trivially obeys Onsager reciprocal relations. However, the CM momentum density does affect the reactive part of the dynamics of the angular momentum density. This, by itself is a manifestation of non-reciprocity. 
%
% odd viscosity appears in non-interacting systems and also odd pressure and 2 other terms 
%
Indeed, when writing the reactive part of the dynamics of the hydrodynamic momentum density, which includes both CM linear momentum density and SAM density, odd viscosity appears together with an odd pressure and two other terms that couple vorticity to shears involving the direction of $\vecl$. We, therefore, conclude that the mere existence of a non-vanishing SAM density breaks Onsager reciprocal relations and gives rise to odd viscosity.
%
% however, no waves propagate even in 3D. In contrast to our PRL. 
%Our new results are verified using kinetic theory.
%
Unlike our previous work~\cite{MarLub2021}, using our direct CG approach we find that there are no new excitations in a 3D fluid (or gas) of non-interacting spinning constituents. Instead, a longitudinal wave will always be accompanied by a transverse wave, but not vice versa, thus breaking Onsager reciprocity. To verify the discrepancy between our two CG approaches we have verified our current results using a kinetic theory.

% Interactions give rise to another odd viscosity that obeys Onsager
%together with \ell this gives rise to propagation of waves and instabilities. Instability lines are exceptional point --> non reciprocal phase transitions
%
It is expected that in sufficiently dense chiral fluids, interactions will play an increasing role. Although we did not propose a microscopic model for such interactions, recent work~\cite{odd_ideal_gas,VitelliKubo} suggests that these exist and give rise to another odd viscosity that obeys Onsager reciprocal relations. Considering such an effect together with the ``kinetic odd viscosity'' gives rise to a non-Hermitian dynamical matrix. Analysing the excitation spectrum of this matrix in 3D reveals regions in which waves propagate (as we found in~\cite{MarLub2021}) and regions that are linearly unstable, suggesting an emergence of a new inhomogeneous phase. Surprisingly, the boundaries that separate these regions are densely packed surfaces of exceptional points, which may indicate the existence of a non-reciprocal phase transition~\cite{Vitelli2021}. As there is no spontaneous symmetry breaking in our system and also because $\vecl$ breaks both chirality and reciprocity simultaneously, our system does not seem to precisely fit within the framework of~\cite{Vitelli2021}. To further study the transition and classify it, one must go beyond the linearization scheme we used to obtain the excitation spectrum. We intend to investigate this in future work.

%our results highlight the siginificance of using the total momentum in chiral active systems
In a broader perspective, our results highlight the significance of using the hydrodynamic momentum density in the study of chiral active materials where angular momentum density does not vanish. 
In passive fluids, the CM and hydrodynamic momentum densities are essentially identical; in chiral active  materials, they differ considerably.  As we have shown, it is the hydrodynamic momentum that is related to actual forces. 
%
%Unlike the case for passive fluids, in such chiral active materials, there is a significant difference between the stress of the CM momentum density and the one related to the total momentum density. As we have shown, the latter is the stress that is related to the actual force on the systems boundaries.
%
Without accounting for the stress caused by the spinning of the molecules one cannot resolve the true forces that acts on the system boundaries.
%
%Importantly, when measuring surface forces in a fluid with spinning particles, these must be determined using the stress tensor of the total momentum density rather than the one related to the CM momentum density. 

\section*{Materials and Methods}

\subsection*{Poisson-Bracket for fields}\label{sec:SI_PB_derivation}

In the main text we show how to derive the hydrodynamic momentum dynamics from the well-known dynamics of the CM and angular momenta. Here we detail the derivation of these equations using the Poisson-bracket (PB) formalism for fields~\cite{Lubensky2003,MarLub2021,HohenbergHal1977,LubenskyBook}. 
%to derive the reactive part of the dynamics for $\vecg$, $\vecl$ and $\rho$. In this section we provide a brief reminder of this method. 
%
We are interested in the dynamics of the coarse-grained total momentum, CM momentum, and angular momentum densities. The microscopic fields $\hat \Phi_\mu\left(\{{\bf q}^\alpha_i\},\{{\bm \pi}^\alpha_i\},t\right)$, with $\{{\bf q}^\alpha_i\}$ and $\{{\bm \pi}^\alpha_i\}$  the generalized coordinates and conjugate momenta, are then coarse-grained to give mesoscopic fields $\Phi_\mu(\vecr,t) = [ \hat \Phi_\mu\left(\{{\bf q}^\alpha_i\},\{{\bm \pi}^\alpha_i\},t\right)]_c$ (the definition of $[\hat{\cal O}]_c$ can be found in another section below), and the statistical mechanics of the latter is determined by the coarse-grained Hamiltonian $H[\{\Phi_\mu\}]$. The reactive (non-dissipative) part of the dynamics of the coarse-grained fields is found by using~\cite{Lubensky2003,MarLub2021,Lubensky2005,Kung2006}
\begin{eqnarray}
\label{p1}
\frac{\partial \Phi_\mu(\vecr,t)}{\partial t} \Big|_{\rm reactive} \!\!\!\!= - \!\!\int \D\vecr' \{\Phi_\mu(\vecr),\Phi_\nu(\vecr')\}\frac{\delta H}{\delta \Phi_\nu(\vecr')}  , \,
\end{eqnarray}
where $\{\Phi_\mu(\vecr),\Phi_\nu(\vecr')\} = [ \{\hat\Phi_\mu(\vecr),\hat\Phi_\nu(\vecr')\}]_c$ and 
\begin{eqnarray}
\label{p1a}
\{\hat\Phi_\mu(\vecr),\hat\Phi_\nu(\vecr')\} \!\!\!\!\!\! &&= \sum_{\alpha i} \Bigg[ \frac{\partial \hat\Phi_\mu(\vecr)}{\partial \pi_i^\alpha} \frac{\partial \hat\Phi_\nu(\vecr')}{\partial q_i^\alpha} \nonumber \\
&&\qquad\qquad - \,\,\frac{\partial \hat\Phi_\mu(\vecr)}{\partial q_i^\alpha} \frac{\partial \hat\Phi_\nu(\vecr')}{\partial \pi_i^\alpha} \Bigg] \, .
\end{eqnarray}
%
%Note that the PB only couples the fields $\Phi_\mu$ and $\Phi_\nu$ if $\Phi_\mu$ have different sign under time reversal than $\delta H / \delta \Phi_\nu$.

Unlike the microscopic dynamics, PBs do not produce the complete mesoscopic dynamics. 
%does not give the complete mesoscopic dynamics. 
The CG procedure neglects many degrees of freedom, which appear as an additional dissipative term in the coarse-grained dynamics
\begin{eqnarray}
\label{p2}
\frac{\partial \Phi_\mu(\vecr,t)}{\partial t} \Big|_{\rm dissipative} = -\int \D\vecr' \, \Gamma_{\mu\nu}(\vecr,\vecr') \frac{\delta H}{\delta \Phi_\nu(\vecr')} \, .
\end{eqnarray}
The dissipative tensor ${\bm \Gamma}$ is symmetric and positive semi definite.
It is generally a function of all fields $\{\Phi_\mu\}$, and following Curie's symmetry principle, it must obey the system symmetries.
Moreover, $\Phi_\mu$ is dissipatively coupled  to $\Phi_\nu$ only if  they have the same sign under time reversal. This is the signature of dissipation where the flux $\partial \Phi_\mu / \partial t$ has the opposite sign under time reversal from its conjugate force $\delta H / \delta \Phi_\nu$.

%For our specific case we use the following fundamental PB (all other PB vanish) and \eqref{p1} to derive the dynamics of the CM momentum: 
%%
%\begin{subequations}
%	\begin{align}
%		\label{p5}
%		&\{\ell_i(\vecr),\ell_j(\vecr')\} = - \varepsilon_{ijk} \ell_k(\vecr) \delta(\vecr-\vecr') \, ,\\
%		\label{p5aa}
%		&\{\ell_i(\vecr),g_j^c(\vecr')\} = \ell_i(\vecr') \nabla_j \delta(\vecr-\vecr') \, ,\\
%		\label{p5a}
%		&\{g^c_i(\vecr),\rho(\vecr')\} = \rho(\vecr)  \nabla_i \delta(\vecr'-\vecr) \, ,\\
%		&\{g_i^c(\vecr),g_j^c(\vecr')\} = g_i^c(\vecr') \nabla_j \delta(\vecr-\vecr') - \nabla'_i \!\left[ g_j^c(\vecr') \delta(\vecr-\vecr')  \right] \, .
%	\end{align}
%\end{subequations}
%%%

\subsection*{Coarse-graining the kinetic energy of complex molecules}

A necessary ingredient in the PB formalism is the knowledge of the coarse-grained Hamiltonian. In this section we use the CG procedure that is detailed in the next section to directly coarse-grain the kinetic energy of a fluid of dumbbells (complete derivation, including for a general complex molecule is deferred to Sec.~III of the SI).
%	
%	
%%%%%%%%%%%% CG the Hamiltonian  %%%%%%%%%%%%%%%
%
%We start by directly coarse-graining the kinetic Hamiltonian:
The microscopic kinetic Hamiltonian is:
\begin{eqnarray}
\label{m1}
H_k =  \sum_{\alpha\mu} \frac{\left(\vecp^{\alpha\mu}\right)^2}{2m^{\alpha\mu}} \, ,
\end{eqnarray}
where the atoms $\mu$ of molecule $\alpha$ are subjected to constraints that force them to move as a rigid body (one can consider also the vibrational motion of the atoms within the molecule, but these finite-frequency modes will not contribute to the hydrodynamics, see Sec. II of the SI).
%fast to their equilibrium values, and we ignore them)}~\footnote{\TM{One can consider also the vibrational motion of the atoms within the molecule, but these will relax fast to their equilibrium values, and we ignore them. A typical example is a diatomic molecule, e.g., ${\rm H}_2$. It has a total of six degrees of freedom: three arising from translation, two from rigid rotation (rotations along the connecting axes are excluded), and one from changes in the separation between the two hydrogen atoms. The latter is a ``fast'' degree of freedom that relaxes rapidly to its equilibrium value.}}. 
%
The first step before CG is to write the kinetic energy density ${\cal H}_k$ such that $H_k = \int \D\vecr \, {\cal H}_k$:  
\begin{eqnarray}
\label{m2}
{\cal H}_k(\vecr) =  \sum_{\alpha\mu} \frac{\left(\vecp^{\alpha\mu}\right)^2}{2m^{\alpha\mu}} \delta\left(\vecr-\vecr^{\alpha\mu}\right) \, .
\end{eqnarray}
There are (at least) three ways of CG the kinetic energy: (i) To ignore all  rigid body constraints and use the CG method described in the next section to obtain ${\cal H}^{({\rm i})}_k(\vecr) = \left[\vecg(\vecr)\right]^2/(2\rho(\vecr))$, where $\vecg$ is the hydrodynamic momentum. This is what we did in Ref.~\cite{MarLub2021}. (ii) Writing the kinetic energy of each rigid molecule using its normal zero-frequency modes and their associated generalized momenta, which are simply the CM and angular momenta~\cite{Goldstein_book}:
\begin{eqnarray}
\label{m7a}
{\cal H}^{({\rm ii})}_k(\vecr) \!=\!  \sum_{\alpha} \left[ \left(\vecp^{\alpha}\right)^2/\left(2M\right) + \left(\vecl^\alpha\right)^2/(2I) \right]\! \delta\left(\vecr-\vecr^{\alpha}\right) \, , \nonumber\\
\end{eqnarray}
%
%${\cal H}_k(\vecr) =  \sum_{\alpha} \left[ \left(\vecp^{\alpha}\right)^2/\left(2M\right) + \left(\vecl^\alpha\right)^2/(2I) \right] \delta\left(\vecr-\vecr^{\alpha}\right)$, 
where the sum is now over the molecules, $\vecp^\alpha$ is the molecule CM momentum with $M$ the molecular mass (for a fluid of dumbbells it is $M=2m$), $\vecl^\alpha$ is the molecule SAM, and $I$ the molecular moment of inertia (for a dumbbell it is constant, $I=Ma^2$). When coarse-grained this gives ${\cal H}^{({\rm ii})}_k(\vecr) = \left[\vecg^c(\vecr)\right]^2/(2\rho(\vecr)) + \left[\vecl(\vecr)\right]^2/(2I)$ as was used in Refs.~\cite{Lubensky2005,NJP,furthauer2012}. (iii) The third way, which is what we use hereafter, is to use the same CG that gives the hydrodynamic momentum equation, \eqref{eq:total_momentum}, which is just taking the long wavelength limit and then using the CG method of the next section. 
%
%Using the CG described in Materials and Methods, and taking the long wavelength limit, the kinetic energy reads (for complete derivation see SI~\cite{Supp})
The kinetic energy then reads (for complete derivation see Sec.~I of the SI)
\begin{eqnarray}
\label{m7}
H_k = \int\D\vecr \left[ \frac{\left(\vecg^c\right)^2}{2\rho} + \frac{\vecl^2}{2I} + \half\nabla \cdot \left(\vecl \times \vecv^c-\vecv^c\cdot{\bf A}\right) \right] \! .
\end{eqnarray}
Importantly, the last term in Eq.~(\ref{m7}) is a boundary term and can therefore be ignored in the bulk. Its origin, is the same as the $\nabla\times\vecl$ term in \eqref{eq:total_momentum}, it comes from molecules that are only partially within the CG volume (see Sec.~I of the SI). Then, the Hamiltonian in terms of the CM momentum assumes the same form as in CG (ii) in the bulk, which is consistent with classical mechanics textbooks~\cite{Goldstein_book}.

The CG procedure described in the next section looses information about internal correlations within the CG volume. Therefore, if such correlations do not vanish (on average) such that they cannot be treated as noise in a fluctuating hydrodynamic theory~\cite{LLstat}, the CG becomes inaccurate. This is precisely what is happening when CG the kinetic energy using method (i) as described above, which gives ${\cal H}^{({\rm i})}_k = \vecg^2/(2\rho)$. The other CGs, (ii) and (iii), only differ in boundary terms, and both considers explicitly the constraints of a rigid body by using the appropriate generalized momenta. Crucially, the difference between the various CG methods is unimportant in {\it passive} fluids, because the neglected correlations, which are related to the SAM, relax in microscopic times. However, when the SAM is driven as in {\it chiral active materials}, these correlations never decay and the distinction between the various CG methods becomes apparent.

Using the PB formalism with the Hamiltonian of \eqref{m7} gives Eqs.~(\ref{e5})-(\ref{eq:cm_dynamics})  with no odd viscosity (in the CM dynamics). Indeed, as we have shown in the main text,  odd viscosity only appears in the hydrodynamic momentum stress.
Intuitively this can be understood by writing the kinetic Hamiltonian in terms of the hydrodynamic momentum with the help of Eqs.~(\ref{eq:total_momentum}) and (\ref{m7}):
\begin{eqnarray}
\label{m9}
H_k = \int\D\vecr \bigg[ \frac{ \vecg^2}{2\rho} + \frac{\vecl^2}{2I} -\vecl\cdot{\bm\omega} 
+ \half\nabla \cdot \left(\vecl \times \vecv -\vecv\cdot{\bf A}\right)  \bigg] \, , \nonumber \\
\end{eqnarray}
where ${\bm \omega} = \half \nabla\times\vecv$ is the rotation vector and terms $\sim (\nabla\vecl)^2$ were neglected. Here the third term $\sim \vecl\cdot{\bm\omega}$ is the one that was responsible for the appearance of odd viscosity in Refs.~\cite{MarLub2021,Banerjee2017}. (The last term remains a boundary term.) 
This result differs from the coarse-grained Hamiltonian we used in Ref.~\cite{MarLub2021} in which the third term is absent. 
%This term is responsible for the appearance of odd viscosity in Refs.~\cite{Banerjee2017,MarLub2021}. 
One can use directly the PB of the hydrodynamic momentum and this Hamiltonian to derive Eqs.~(\ref{eq:NSE_total_momentum})-(\ref{eq:odd viscosity}). Importantly, this Hamiltonian is written in terms of the hydrodynamic momentum and the SAM, which are {\it dependent} via \eqref{eq:total_momentum}. The two independent fields here are the CM momentum and the SAM, which are coarse-grained from the independent molecular degrees-of-freedom.
%Below we show that odd viscosity emerges from \eqref{m9} as well, in addition to other non-dissipative terms.
%

Note that although the last term in Eqs.~(\ref{m7}-\ref{m9}) can be ignored in the bulk it will affect the boundary conditions and may play a significant role in the study of surface states in fluids of rotating particles~\cite{Abanov2018,Irvine2019,Souslov2019}.

\subsection*{Coarse-graining according to Paul C. Martin}\label{app:cg}

%The PB formalism requires the knowledge of the coarse-grained Hamiltonian. 
In this section we explain in detail the CG procedure we refer to throughout this paper.
Since PB are strictly mechanistic, they do not require any ensemble average or the introduction of temperature. Here, we instead perform a strictly mechanical average. 
%Coarse-graining is used quite loosely in the literature, sometimes it is referring to smoothing out functions (homogenization??)~\cite{XXX} and sometimes to averaging over some volume $\Delta V$ which is large microscopically but small macroscopically~\cite{XXX}. 
In most of literature smoothing  $\hat{\boldmath{\cal O}} = \sum_\alpha \boldmath{\cal O}^\alpha\delta(\vecr-\vecr^\alpha)$ is done as follows (see, e.g.,~\cite{klymko2017}):
\begin{eqnarray}
\nonumber\boldmath{\cal O}(\vecr) \equiv \left[\hat{\boldmath{\cal O}}(\vecr)\right]_c  &=& \int\D\vecr' W(\vecr-\vecr') \hat{\boldmath{\cal O}}(\vecr')  \\
&=& \sum_\alpha \boldmath{\cal O}^\alpha W(\vecr-\vecr^\alpha) \, , \label{s1}
\end{eqnarray}
where $W$ is a smooth function such as a Gaussian or a Heaviside function that obeys $\int\D\vecr W(\vecr)=1$. 

\begin{figure}
\begin{centering}
\includegraphics[width=0.3\textwidth]{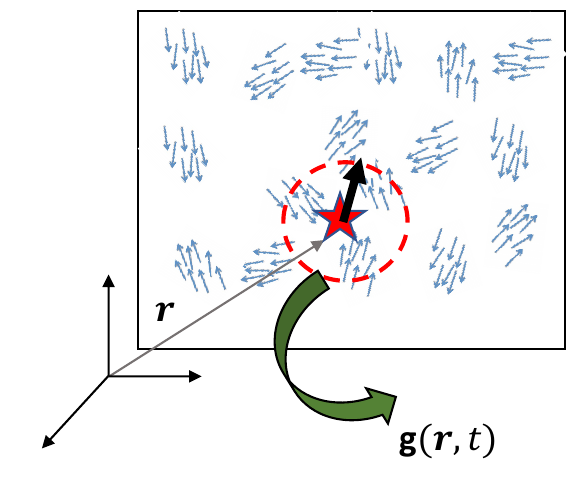}
\par\end{centering}
\caption{ Illustration of coarse-graining the momentum density at point $\vecr$ in time $t$. Small arrows are the momentum of various molecules in the system. The large thick arrow shows the average momentum within the CG volume, which is marked by the red dashed circle.
}
\label{fig:CG} 
\end{figure}
%

%
%Usually such mechanical CG, which is sometimes refer to as homogenization~\cite{XXX}, aims at smoothing a function $\hat{\boldmath{\cal O}}$ by averaging over some volume in space. In most of literature smoothing  $\hat{\boldmath{\cal O}} = \sum_\alpha \boldmath{\cal O}^\alpha\delta(\vecr-\vecr^\alpha)$ is done as follows:
%
%\begin{eqnarray}
%\label{s1}
%\nonumber\boldmath{\cal O}(\vecr) \equiv \left[\hat{\boldmath{\cal O}}(\vecr)\right]_c  &=& \int\D\vecr' W(\vecr-\vecr') \hat{\boldmath{\cal O}}(\vecr') \\
%&=& \sum_\alpha \boldmath{\cal O}^\alpha W(\vecr-\vecr^\alpha)\, ,
%\end{eqnarray}
%
%where $W$ is a smooth function such as a Gaussian or a Heaviside function that obeys $\int\D\vecr W(\vecr)=1$. 
%

We, however, are not interested in this homogenization procedure (which we refer to as usual), but rather in performing spatial averages within cells of volume $\Delta V$ (defined by the width of $W$), see Fig.~\ref{fig:CG}. In this process of averaging we disregard any other intra-cell information. This extra intra-cell information, is the cause of (thermal) fluctuations, which can be added to our hydrodynamic equations~\cite{LLstat}. We discuss this briefly in the end of the kinetic theory derivation below. %Sec.~\ref{sec:kinetic}. 
Our aim is to approximate, within such CG volume, functions of the form:
\begin{eqnarray}
\label{e1}
{\boldmath\hat{\cal O}}(\vecr) = \sum_\alpha \boldsymbol{A}^\alpha \boldsymbol{B}^\alpha \delta(\vecr-\vecr^\alpha) \, ,
\end{eqnarray}
using the coarse-grained fields ${\boldsymbol{A}}(\vecr) = \sum_\alpha \boldsymbol{A}^\alpha W(\vecr-\vecr^\alpha)$ and ${\boldsymbol{B}}(\vecr) = \sum_\alpha \boldsymbol{B}^\alpha W(\vecr-\vecr^\alpha)$. To do so we follow Paul C. Martin ideas and replace $\boldsymbol{B}^\alpha$ (or $\boldsymbol{A}^\alpha$) with its average within the CG volume, $\bar{\boldsymbol{B}}(\vecr) =  \boldsymbol{B}(\vecr)/n(\vecr)$, where $n(\vecr) \equiv \sum_\alpha  W(\vecr-\vecr^\alpha)$. 
%%
%\begin{eqnarray}
%\label{e1a}
%\bar{\boldsymbol{B}}(\vecr) =  \boldsymbol{B}(\vecr)/n(\vecr)\, ,
%\end{eqnarray}
%%
%$\bar{\boldsymbol{B}}(\vecr) =  \boldsymbol{B}(\vecr)/n(\vecr)$, 
%where  $\rho(\vecr) \equiv \sum_\alpha m^\alpha W(\vecr-\vecr^\alpha)$ and we define the avreage mass per particle via $\rho(\vecr)=m n(\vecr)$ with  
%where $n(\vecr) \equiv \sum_\alpha  W(\vecr-\vecr^\alpha)$. 
%
%If we choose $W(\vecr)$ to be a Heaviside function we have
%%
%\begin{eqnarray}
%\label{e2}
%\nonumber\bar{\boldsymbol{B}}(\vecr) &=& \frac{\int\D\boldsymbol{u} \sum_{\alpha} \boldsymbol{B}^\alpha \delta(\vecr-\boldsymbol{u}-\vecr^\alpha)} 
%{\int\D\boldsymbol{u} \sum_\alpha \delta(\vecr-\boldsymbol{u}-\vecr^\alpha)}  \\
%\nonumber&=& \frac{1}{\Delta V\hat{n}(\vecr)} \int\D\boldsymbol{u} \sum_{\alpha} \boldsymbol{B}^\alpha \delta(\vecr-\boldsymbol{u}-\vecr^\alpha)  \\
%&\simeq& \frac{\hat{\boldsymbol{B}}(\vecr)}{\hat{n}(\vecr)}\, ,
%\end{eqnarray}
%%
%where in the last step we have assumed the CG volume is small such that the $\int_{\Delta V} f(\vecr-\vecr')\D\vecr' \simeq \Delta V f(\vecr)$, with $\Delta V$ being the CG volume. %We also define here the number density $n(\vecr) \equiv \sum_\alpha W(\vecr-\vecr^\alpha)$. 
With this we can write,
\begin{eqnarray}
\label{e3}
{\boldmath{\cal O}}(\vecr) \simeq \bar{\boldsymbol{B}}(\vecr) \sum_{\alpha} \boldsymbol{A}^\alpha W(\vecr-\vecr^\alpha) 
= \frac{\boldsymbol{A}(\vecr)\boldsymbol{B}(\vecr)}{n(\vecr) }\, .
\end{eqnarray}
In Sec.~VI of the SI we provide simple examples and some extensions of this method. 

Interestingly, if one chooses cells that are of the size of a dumbbell (or a rigid molecule in the general case), such that only one dumbbell can be within each cell, an exact relation for the microscopic fields ${\boldmath\hat{\cal O}}(\vecr)$ can be obtained~\cite{Nakamura2009}:
\begin{eqnarray}
\nonumber{\boldmath\hat{\cal O}}(\vecr) &=& \sum_\alpha \boldsymbol{A}^\alpha \boldsymbol{B}^\alpha \delta(\vecr-\vecr^\alpha) 
\frac{\sum_\beta \delta(\vecr-\vecr^\beta)}{\sum_\gamma \delta(\vecr-\vecr^\gamma)} \\
\nonumber&=& \frac{1}{\hat{n}(\vecr)} \sum_{\alpha,\beta} \boldsymbol{A}^\alpha \boldsymbol{B}^\beta \delta(\vecr-\vecr^\alpha) \delta(\vecr-\vecr^\beta) \\
&=&\frac{\hat{\boldsymbol{A}}(\vecr) \hat{\boldsymbol{B}(\vecr)}}{\hat{n}(\vecr)} \, ,  \label{e8}
\end{eqnarray}
where $\hat{\cal O}(\vecr)$ denotes a microscopic field. The second line of \eqref{e8} is exact because no two molecules occupy the same cell. In fact, using this equation we can write the Hamiltonian of \eqref{m7} in terms of the microscopic fields and calculate the microscopic PBs, thus getting the reactive part of the dynamics in terms of the microscopic fields. These microscopic dynamics assumes exactly the same form as Eqs.~(\ref{e5})-(\ref{eq:total_momntum_dynamics}) (but with the microscopic fields $\{\hat{\cal O}_i\}$), such that odd viscosity is present even in this microscopic formulation. In the microscopic formulation there is, however, no dissipation (in contrast to \eqref{eq:NSE_total_momentum}) and $F[\rho]$ of \eqref{eq:H_tot} can no longer be thought of as free energy and it must be written in terms of the microscopic fields (see, e.g., $F$ in the kinetic theory derivation below).

Note that in Ref.~\cite{Nakamura2009} a single atom is allowed within a cell, while here a single complex molecule  (dumbbell in the simplest case) can be within a cell, which leads to the appearance of the odd terms that are related to the molecule angular momentum.
If one insists on CG using cells that allow occupation of a single atom, the rigid body constraints creates correlated motion between near coarse-grained volumes. This is an undesired artifact that requires the addition of forces that will maintain such constraints. Treating this is much more complicated than considering the molecules as rigid bodies and coarse-grain using cells containing single molecules as we described above.
In this paper we are dealing with CG of a fluid of complex molecules, where we find that there is a difference between the stress associated with the total momentum density and the one associated with the CM momentum density.
Obviously, the total momentum coincide with the CM momentum for point-like particles, which is related to the fact that point-like particles cannot rotate and thus do not posses angular momentum.  A more complex molecule or ``shaped particle'' is needed to support angular momentum. The simplest type of such molecule is a dumbbell, which is what we focus on in the main text.  
As noted above, in the process of CG we disregard any deviation from the average of any field within the coarse-grained volume. When considering complex molecules, this includes any vibrational modes of the molecules, such that only the molecules zero-modes contribute to the coarse-grained fields. Intuitively, the culling of finite-frequency modes is a result of averaging over length-scales much larger than the vibration amplitude (long wave-length) and over time-scales much longer than the vibration period (low frequency limit). Each of these averages on its own will remove any finite-frequency modes (in the SI Secs.~I-II we show both of these processes independently). It is, therefore, important to identify the zero-modes of the molecules before CG.

\section*{Dynamics of the hydrodynamic momentum}

In terms of the CM momentum, the kinetic Hamiltonian of \eqref{m7} is quite standard (disregarding the boundary term) and the complete Hamiltonian is~\cite{MarLub2021,Lubensky2003}:
\begin{eqnarray}
\label{eq:H_tot}
H = \int \D \vecr  \left( \frac{\left(\vecg^c\right)^2}{2\rho} + \frac{\vecl^2}{2I} + F[\rho(\vecr)] \right)   \, ,
\end{eqnarray}
where $F[\rho] = \int\D\vecr {\cal F}\left(\rho,\nabla\rho\right)$ is the free energy, which in the isotropic case is only a functional of $\rho(\vecr)$. The latter statement assumes that interactions are only central-force, i.e., they only depend on the CM positions of the molecules. When friction between molecules is considered this might not be the case and other types of interactions that depend on $\vecl$ may appear~\cite{VitelliKubo} -- we ignore such interactions for now.

To derive the dynamics of the coarse-grained fields we use the PB formalism for fields~\cite{Lubensky2003,MarLub2021,Kung2006} (see section above). For our specific case we use the following fundamental PB (all other PB vanish): 
\begin{subequations}
\begin{align}
\label{p5}
&\{\ell_i(\vecr),\ell_j(\vecr')\} = - \varepsilon_{ijk} \ell_k(\vecr) \delta(\vecr-\vecr') \, ,\\
\label{p5aa}
&\{\ell_i(\vecr),g_j^c(\vecr')\} = \ell_i(\vecr') \nabla_j \delta(\vecr-\vecr') \, ,\\
\label{p5a}
&\{g^c_i(\vecr),\rho(\vecr')\} = \rho(\vecr)  \nabla_i \delta(\vecr'-\vecr) \, ,\\
\nonumber&\{g_i^c(\vecr),g_j^c(\vecr')\} = g_i^c(\vecr') \nabla_j \delta(\vecr-\vecr') \\
&\qquad\qquad\qquad\qquad- \nabla'_i \!\left[ g_j^c(\vecr') \delta(\vecr-\vecr')  \right] \, .
\end{align}
\end{subequations}
Substituting these PB into \eqref{p1} and using the Hamiltonian of \eqref{eq:H_tot} we derive the reactive part of the dynamics for $\vecg^c$, $\vecl$ and $\rho$:
%to derive the dynamics of the CM momentum
%
\begin{subequations}
\begin{align}
\label{dtm1}
&\dot\rho + \nabla\cdot\vecg^c = 0 \, , \\
\label{dtm2}
&\dot{g}_i^c + \nabla_j \left(v_j^c g_i^c\right) = -\nabla_i P + f_i \, ,\\
\label{dtm3}
&\dot{\ell}_i + \nabla_j \left(v_j^c \ell_i\right) = \tau_i \, , 
\end{align}
\end{subequations}
where $\boldsymbol{f}$ and ${\bm \tau}$ are, respectively, external force and torque densities that were added to the dynamics of linear and angular momentum. Here $P = \rho\frac{\delta F}{\delta\rho} - {\cal F}$ is the thermodynamic pressure and $F=\int\D\vecr{\cal F}$. This result is nor surprising or interesting, we simply recovered the continuity equation, the reactive parts of the Navier-Stokes equation and the angular momentum density dynamics. When adding the usual dissipative terms for isotropic fluids we obtain Eqs.~(\ref{e5})-(\ref{eq:cm_dynamics}).
It is also possible to derive directly \eqref{eq:total_momntum_dynamics} using the PB formalism and the Hamiltonian of \eqref{m9}.

\subsection*{Kinetic theory derivation}\label{sec:kinetic}

The kinetic derivation is essentially the same as the Irving-Kirkwood derivation of the equations of hydrodynamics~\cite{Irving-Kirkwood}, which uses the Liouville equation for the probability density and ensemble average. 
%It can be shown (see SI~\cite{Supp}) that these two approaches are essentially the same, where the only difference lies in the type of averaging. In the Irving-Kirkwood approach an ensemble average is used, while in the kinetic theory derivation~\cite{klymko2017} an average over some small volume (coarse-graining) is used.
The only difference between the approaches lies in the type of averaging. In the Irving-Kirkwood approach an ensemble average is used, while in the kinetic theory derivation~\cite{klymko2017} an average over some small volume (coarse-graining) is used. If the coarse-grained volume contains a large number of molecules these two averages coincide.

We start from the microscopic equation for the hydrodynamic momentum density (see \eqref{eq:g_s}) and ignore the $\dot{\bm Q}$ term that will vanish once the equation is coarse-grained:
\begin{eqnarray}
\label{k1}
\hat{\vecg}(\vecr) \simeq \sum_{\alpha} \left[ \vecp^{\alpha} + \half\nabla\times\vecl^\alpha  \right] \delta(\vecr-\vecr^{\alpha}) \, ,
\end{eqnarray}
where as before $\vecp^{\alpha} = \sum_\mu \vecp^{\alpha\mu}$ and $\vecl^\alpha = I {\bm \nu}^\alpha \times \dot{\bm\nu}^\alpha$. We remind the reader that, similarly to the main text, by writing \eqref{k1} we coarse-grain the atoms forming the molecule and treat every molecule as a point-like particle with SAM. Assuming the molecules  only interact via two-body central forces (this is of course not general -- we expect interactions involving mutual torques will lead to another odd viscosity term on their own~\cite{VitelliKubo}) we have $\dot{\vecp}^{c,\alpha} = -\nabla_\alpha U\left(\{\vecr^\beta\}\right) + \boldsymbol{f}^\alpha$ and $\dot{\vecl}^\alpha = {\bm T}^\alpha$ such that: 
\begin{eqnarray}
\nonumber\partial_t \hat{g}_i &=& \sum_\alpha \bigg[ f_i^\alpha -\nabla_\alpha U\left(\{\vecr^\beta\}\right) -p^\alpha_i v_j^\alpha \nabla_j  \\
&-& \half\varepsilon_{ikn}\ell^\alpha_n v_j^\alpha \nabla_j \nabla_k + \half\varepsilon_{ijk} T^\alpha_k \nabla_j \bigg] \delta(\vecr-\vecr^{\alpha}) \, .  \quad
\label{k2}
\end{eqnarray}
Here $U\left(\{\vecr^\beta\}\right) = \half \sum_{\alpha\neq \beta} \phi\left(\left| \vecr^\alpha-\vecr^\beta \right|\right)$ is a general two-body potential.
Coarse-graining by employing \eqref{e3} and using \eqref{eq:total_momentum} we get
\begin{eqnarray}
& &\nonumber \dot{g}_i + \nabla_j(g_i v_j) = f_i -\nabla_i P \\
& &+\nabla_j \left[ - \half \ell_n \left( \varepsilon_{iln}\delta_{jk} + \varepsilon_{jln}\delta_{ik}\right)   \nabla_l v_k + \half\varepsilon_{ijk} \tau_k  \right] \, ,  \quad\quad
\label{k6}
\end{eqnarray}
with $\boldsymbol{f} \equiv \sum_\alpha \boldsymbol{f}^\alpha  \delta(\vecr-\vecr^{\alpha})$ and ${\bm \tau}(\vecr) \equiv \sum_\alpha \boldsymbol{T}^\alpha  \delta(\vecr-\vecr^{\alpha})$ being the external force and torque densities, respectively, and $F=\int\D\vecr\D\vecr' \left[ \rho(\vecr) \phi\left(\left|\vecr-\vecr'\right|\right) \rho(\vecr')\right]$.
The pressure is as in the main text, $P = \rho\frac{\delta F}{\delta\rho} - {\cal F}$, and we have dropped a non hydrodynamic term $\sim \nabla \left(\nabla\times\vecl\right)^2$. This result can be shown to be equivalent to that obtained in the main text  from the PB approach for the standard kinetic energy (see Sec.~IV of the SI).

Note that within the kinetic derivation above, the thermal pressure term is absent (the pressure above is a result of molecule interactions). In some literature (e.g., \cite{klymko2017}) the kinetic part of the stress (which is the thermal pressure) is written in a way similar to that of the Irving-Kirkwood kinetic stress~\cite{Irving-Kirkwood}, $\sigma^K_{ij} = - \sum_\alpha m^\alpha \left(\frac{\vecp^\alpha}{m^\alpha} - \vecv\right)_i \left(\frac{\vecp^\alpha}{m^\alpha} - \vecv\right)_j  W(\vecr-\vecr^\alpha)$. We use the fluctuating hydrodynamics approach~\cite{LLstat,HohenbergHal1977} in which any deviation from the average motion appears as thermal noise. In such an approach, ${\bm \sigma}^K$ is absent before any ensemble average. It does appear when considering the ensemble average of the dynamics. By writing $\vecv = \langle \vecv \rangle + \delta\vecv$ and $\rho = \langle\rho\rangle +\delta\rho$ (where $\langle ... \rangle$ denotes average over noise realizations), the streaming term becomes $\nabla_j (g_i v_j) = \nabla_j \left( \langle \rho \rangle \langle v_i \rangle \langle v_j \rangle + \langle \rho \rangle \langle \delta v_i \delta v_j \rangle \right)$, where the second term gives the ideal gas pressure as desired.  
Note also that by adding the ideal gas entropy $-T\int\D\vecr \left[ n\log n - n \right]$  to $F$ (thus making it a free-energy rather then just internal energy) the dynamics obtained (without the addition of noise) are the same as after the thermal averaging discussed above.

\begin{acknowledgments} 
	\emph{Acknowledgments:} 
	We thank an anonymous referee for helpful comments and suggestions. This research was supported in part by Grant No.2022/369 from the United States-Israel Binational Science Foundation (BSF). T.M. acknowledges funding from the Israel Science Foundation (Grant No.1356/22). T.C.L. acknowledges funding from the NSF Materials Research Scienceand Engineering Center (MRSEC) at University of Pennsylvania (Grant No.DMR-1720530).
\end{acknowledgments}

% Bibliography

%\bibliography{citation.bib}

%apsrev4-2.bst 2019-01-14 (MD) hand-edited version of apsrev4-1.bst
%Control: key (0)
%Control: author (8) initials jnrlst
%Control: editor formatted (1) identically to author
%Control: production of article title (0) allowed
%Control: page (0) single
%Control: year (1) truncated
%Control: production of eprint (0) enabled
%

\end{document}

%\title{Supplementary Material \\  Non reciprocal odd viscosity: Coarse graining the kinetic energy and exceptional instability}
%\title{Supplementary Material \\ \TM{Chiral active matter is non reciprocal: Implications on odd viscosity and exceptional instability}}
\title{Supporting Information:\\ Non reciprocity and odd viscosity in chiral active fluids}

\author{Tomer Markovich$^{1,2}$}
\email{tmarkovich@tauex.tau.ac.il} 
\author{Tom C. Lubensky$^{3}$}
\affiliation{
	$^{1}$School of Mechanical Engineering, Tel Aviv University, Tel Aviv 69978, Israel \\
	$^{2}$Center for Physics and Chemistry of Living Systems, Tel Aviv University, Tel Aviv 69978, Israel \\
	$^{3}$Department of Physics and Astronomy, University of Pennsylvania, Philadelphia, Pennsylvania 19104, USA
}

\date{\today}

\maketitle

\section{Total hydrodynamic momentum and kinetic energy of a fluid of dumbbells} \label{sec:diatomic}

In this section we detail the derivation of Eq.~(1) of the main text, and Eq.~(18) in Materials and Methods. 
Considering a fluid of dumbbells as described in the main text, the momentum of the two point masses can we written in terms of the CM momentum, $\vecp^\alpha$, and the unit vector ${\bm \nu}^\alpha$ as $\vecp^{\alpha}_{1,2} = \half\vecp^\alpha \pm ma\dot{\bm\nu}^\alpha$,
such that the total hydrodynamic momentum density for the diatomic fluid is
\begin{eqnarray}
\label{eq:total_momentum1}
\nonumber\hat{g}_i(\vecr) &=& \sum_{\alpha\mu} p_i^{\alpha\mu} \delta \left(\vecr - \vecr^{\alpha\mu}\right) \\
\nonumber&=&\sum_\alpha \left[ p_i^{\alpha,1} \delta(\vecr-\vecr^\alpha - a{\bm \nu}^\alpha) + p_i^{\alpha,2}  \delta(\vecr-\vecr^\alpha + a{\bm \nu}^\alpha)   \right] \\
&\simeq&  \sum_\alpha \left[p_i^\alpha \delta(\vecr-\vecr^\alpha) - I\dot{\nu}_i^\alpha \nu_j^\alpha \nabla_j  \delta(\vecr-\vecr^\alpha)  \right] \, ,
\end{eqnarray}
where we have used $\vecr^{\alpha,1} - \vecr^{\alpha,2} = 2a{\bm \nu}^\alpha$, and the moment of inertia of each molecule is $I=Ma^2$ with $M=2m$ being the molecule mass.
As we are interested in the long wavelength limit, $a \ll |\vecr-\vecr^\alpha|$, we only keep terms ${\cal O}(a^2)$.
It is useful to write $\hat\vecg = \hat\vecg^c + \hat\vecg^s$, with $\hat\vecg^c \equiv \sum_\alpha \vecp^\alpha \delta(\vecr-\vecr^\alpha)$ and
\begin{eqnarray}
\label{eq:g_s}
\hat{g}_i^s(\vecr) = -\frac{I}{2} \Big[ \varepsilon_{ijk} \varepsilon_{klm} \sum_\alpha  \dot{\nu}_l^\alpha\nu_m^\alpha \nabla_j\delta(\vecr-\vecr^\alpha) 
+\sum_\alpha \dot{Q}_{ij}^\alpha  \nabla_j\delta(\vecr-\vecr^\alpha)   \Big] \, .
\end{eqnarray}
Here $Q_{ij}^\alpha = \nu_i^\alpha \nu_j^\alpha - \delta_{ij}/d$ is the alignment tensor and $d$ is the number of dimensions. The first term of Eq.~(\ref{eq:g_s}) includes the angular momentum of each molecule, $\vecl^\alpha = I {\bm \nu}^\alpha \times \dot{\bm \nu}^\alpha$. By using the definition of the angular velocity of each molecule, $\Omega^\alpha \equiv {\bm \nu}^\alpha \times \dot{\bm \nu}^\alpha$, one gets $\vecl^\alpha = I {\bm \Omega}^\alpha$. This is Eq.~(1) of the main text.

\begin{figure}
	\begin{centering}
		\includegraphics[scale=0.4]{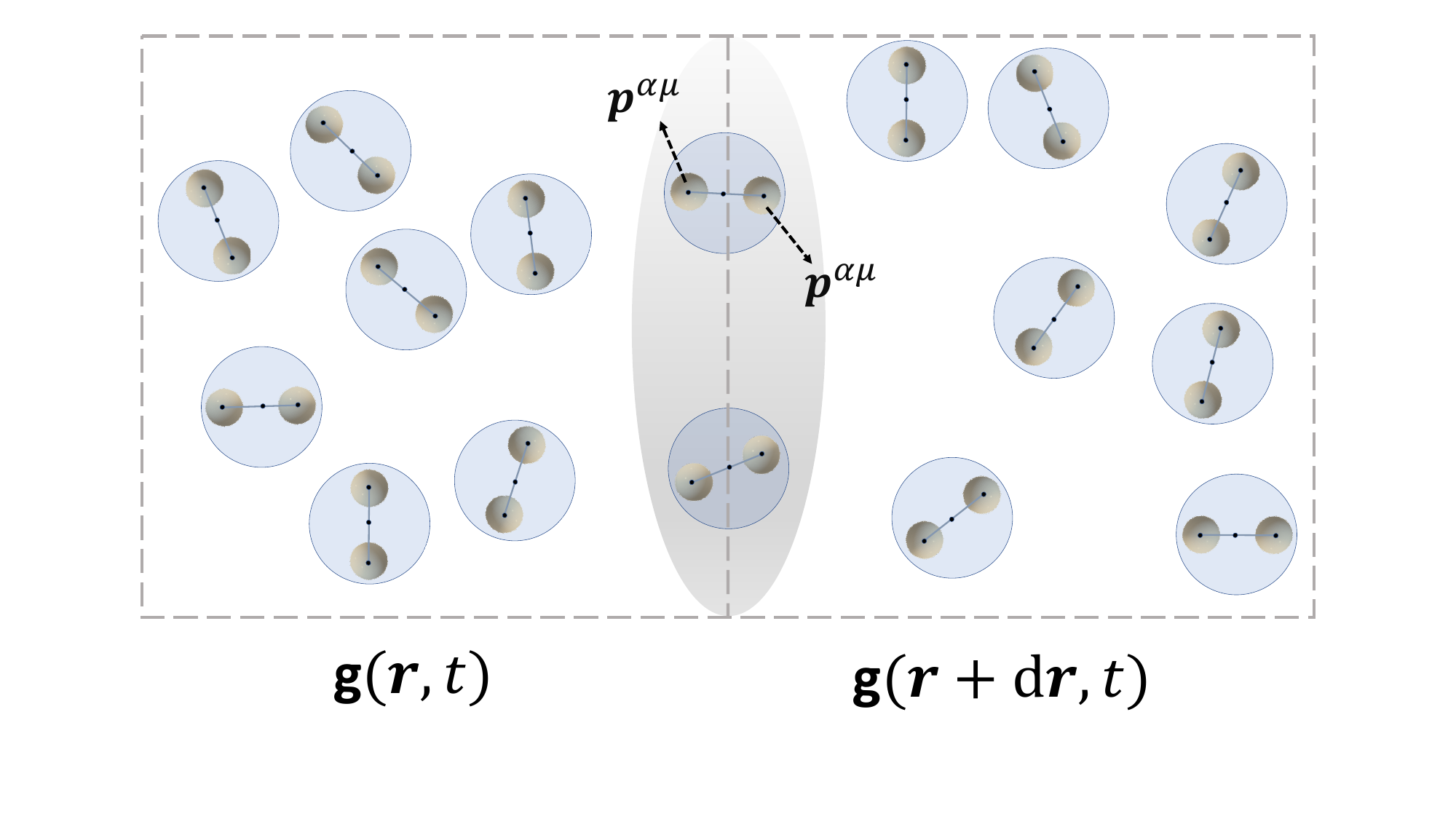}
		\par\end{centering}
	\caption{Cartoon of two adjacent volume elements of a fluid of rotating dumbbells. 
			When a dumbbell is fully within the volume element $\Delta V$ its contribution to the total hydrodynamic momentum of the volume element (the coarse-grained total momentum) is just its CM momentum. However, when a dumbbell is located at the boundary of $\Delta V$ such that one atom is within $\Delta V$ and one is in the adjacent volume element, the rotation of the atoms around the dumbbell CM contributes to the total hydrodynamic momentum within $\Delta V$.
		}
	\label{fig:CG_total_momentum} 
\end{figure}

Note that this equation (which is very similar to the one used in Ref.~\cite{Martin}), although written in terms of the microscopic fields, is already coarse-grained in a way. It accounts for the atoms positions within the molecule only in the long wavelength limit, but does not neglect it completely. We, therefore, refer to $\hat\vecg = \hat\vecg^c + \hat\vecg^s$ as the {\it hydrodynamic momentum}.
This is very similar to the classical treatment of rigid bodies~\cite{Goldstein_book}, where the fast degrees-of-freedom are neglected and a rigid body is modeled as a point-like particle with CM and angular momenta (about the CM). Unlike the classical treatment, by maintaining some information about the atoms position, our coarse-graining (CG) of the total hydrodynamic momentum differ from the CM momentum.
As this difference involves a gradient it is clearly related to a surface contribution. Indeed, its origin is in the fact that a molecule can be part within one CG volume and part in another, such that the angular momentum of the molecule will contribute to the total hydrodynamic momentum within the coarse-grained volume, see Fig.~\ref{fig:CG_total_momentum}.

% such that a molecule can be part within one CG volume and part in another.

%
%

%%%%%%%%%%%% CG the Hamiltonian  %%%%%%%%%%%%%%%

We continue by directly coarse-graining the kinetic Hamiltonian:
\begin{eqnarray}
\label{m1}
H_k =  \sum_{\alpha\mu} \frac{\left(\vecp^{\alpha\mu}\right)^2}{2m^{\alpha\mu}} \, .
\end{eqnarray}
Defining the kinetic energy density ${\cal H}_k$ such that $H_k = \int \D\vecr {\cal H}_k$ we have
\begin{eqnarray}
\label{eq:hamiltonian}
\nonumber {\cal H}_k(\vecr) &=&  \frac{1}{2m}\sum_\alpha \bigg[ \left(\half\vecp^\alpha + ma\dot{\bm\nu}^\alpha\right)^2 \delta\left(\vecr-\vecr^{\alpha} - a{\bm \nu}^\alpha\right) 
 + \left(\half\vecp^\alpha - ma\dot{\bm\nu}^\alpha\right)^2 \delta\left(\vecr-\vecr^{\alpha} + a{\bm \nu}^\alpha\right)   \bigg] \\
&\simeq& \sum_\alpha \left[ \frac{ \vecp_\alpha^2 }{2M} + \half I \dot{\bm \nu}_\alpha^2 
- \frac{I}{M} \left(\dot{\bm \nu}^\alpha\cdot\vecp^\alpha\right) \left( {\bm \nu}^\alpha \cdot \nabla\right)    \right] \delta\left(\vecr-\vecr^{\alpha}\right) \, .,
\end{eqnarray}
where in the second line we have used the long-wavelength limit, which as explained above is already doing some kind of CG. This CG is consistent with the derivation of Eq.~(2) of the main text and Ref.~\cite{Martin}.
We now use the CG method described in Materials and Methods (which coarse-grain many molecules within some volume) to write Eq.~\eqref{eq:hamiltonian} in terms of the various fields. Let us start with the first term of  Eq.~\eqref{eq:hamiltonian}, which after coarse-graining is written as:
\begin{eqnarray}
\label{m3}
\sum_\alpha  \frac{ \vecp_\alpha^2 }{2M} \delta\left(\vecr-\vecr^{\alpha}\right) = \frac{\left(\vecg^c\right)^2}{2\rho} \, ,
\end{eqnarray}
where $\rho(\vecr) = M n(\vecr)$ and $\hat{n}(\vecr) = \sum_\alpha \delta(\vecr-\vecr^\alpha)$. Noting that the relations ${\bm \Omega}^\alpha = {\bm \nu}^\alpha \times \dot{\bm \nu}^\alpha$
and $\dot{\bm \nu}^\alpha \cdot {\bm \nu}^\alpha = 0$ implies that $\left(\dot{\bm \nu}^\alpha\right)^2 = \left({\bm \Omega}^\alpha\right)^2$, coarse-graining the second term of Eq.~\eqref{eq:hamiltonian} yields
\begin{eqnarray}
\label{m4}
\sum_\alpha  \half I \dot{\bm \nu}_\alpha^2 \delta(\vecr-\vecr^\alpha) = \sum_\alpha  \frac{ \left(\vecl^\alpha\right)^2}{2I} \delta(\vecr-\vecr^\alpha)
= \frac{ \vecl^2}{2\tilde{I}\rho}  \, .
\end{eqnarray}
The third term of Eq.~\eqref{eq:hamiltonian} can be written as
\begin{eqnarray}
\label{m5}
\nonumber- \nabla_j \sum_\alpha \tilde{I} \dot{ \nu}^\alpha_i  {\nu}^\alpha_j p^\alpha_i    \delta\left(\vecr-\vecr^{\alpha}\right) 
&=& -\half  \nabla_j \sum_\alpha \tilde{I}  \Big[ \left( \dot{ \nu}^\alpha_i  {\nu}^\alpha_j + \dot{ \nu}^\alpha_j  {\nu}^\alpha_i \right) 
+\varepsilon_{ijk}\varepsilon_{kmn} \dot{ \nu}^\alpha_m  {\nu}^\alpha_n \Big] p^\alpha_i    \delta\left(\vecr-\vecr^{\alpha}\right) \\
&=& -\half  \nabla_j \sum_\alpha \left( \tilde{I} \dot{Q}^\alpha_{ij} - \frac{1}{M}\varepsilon_{ijk} \ell^\alpha_k  \right) p^\alpha_i    \delta\left(\vecr-\vecr^{\alpha}\right) \, ,
\end{eqnarray}
which after coarse-graining reads
\begin{eqnarray}
\label{m6}
- \nabla_j  \sum_\alpha  \tilde{I} \dot{ \nu}^\alpha_i  {\nu}^\alpha_j p^\alpha_i    \delta\left(\vecr-\vecr^{\alpha}\right) 
= \half\nabla \cdot \left(\vecl \times \vecv - \vecv\cdot{\bf A} \right)   , \,\,
\end{eqnarray}
where ${\bf A}(\vecr) = I\sum_\alpha \dot{\bf Q}^\alpha \delta\left(\vecr-\vecr^\alpha\right)$.
Collecting the above, we find that the kinetic energy is the one of Eq.~(18) of Materials and Methods:
\begin{eqnarray}
\label{m7}
H_k = \int\D\vecr \left[ \frac{\left(\vecg^c\right)^2}{2\rho} + \frac{\vecl^2}{2I} + \half\nabla \cdot \left(\vecl \times \vecv -\vecv\cdot{\bf A}\right) \right] \, .
\end{eqnarray}

\section{Fast modes in a dumbbell} \label{sec:fast_modes}

This section will show one approach to the culling of fast modes.  To keep matters simple, we will use our dumbbell model  in two-dimensions modified to have a fast mode. We begin by providing a description of the dynamics of the individual atoms of the dumbbell (total of four dynamical variables) that describe all motions of the dumbbell including stretching and compressing the dumbbell along its axes.  Let ${\bf R}_1$ be the position atom $1$ and ${\bf R}_2$ the position of atom $2$.  Then
%%%
\begin{eqnarray}
\nonumber	{\bf R}_1 &=& {\bf R}_c + (a + r) {\bm\nu}    \, , \\
					   {\bf R}_2 &=& {\bf R}_c - (a+r) {\bm\nu}  \, ,
\label{rvab}
\end{eqnarray} 
%%%
where ${\bf R}_c$ is the center of mass, $a$ is the rest length of the intra-molecular potential $U( \left| {\bf R}_1 - {\bf R}_2\right| )$  %connecting spring, 
and $r$ is the deviation from the rest length. % length of the stretch of the spring.  
From this we calculate the time derivatives:
%%%
\begin{eqnarray}
	\dot{{\bf R}}_1 &=& \dot{{\bf R}}_c + \dot{r} {\bm\nu} + (a + r) \dot{{\bm\nu} } \, ,\nonumber \\
	\dot{{\bf R}}_2 &=& \dot{{\bf R}}_c - \dot{r} {\bm\nu}  - (a + r) \dot{{\bm\nu} } \, .
\end{eqnarray} 
%%%%
%where the last $r$ in both equations can be dropped because it can be assumed to be %we assume it is 
%much smaller than $a$.  
%
The full kinetic energy is then:
%%%
\begin{eqnarray}
	H_k &=& \frac{1}{2} m \left[\dot{{\bf R}}_1^2 + \dot{{\bf R}}_2^2 \right]  \nonumber \\
	&=& \frac{1}{2} M \left[\dot{{\bf R}}_c^2 + \dot{r}^2 + \left(a+r\right)^2 \Omega^2\right]  \, ,
\end{eqnarray} 
%%%%
where  $M=2m$ is the dumbbell mass, and ${\bm\nu} \cdot \dot{\bm\nu}=0$ implies that ${\Omega}^2 = \dot{\bm\nu}^2$.
We have replaced the  four  dynamical variables ${\bf R}_1$ and  ${\bf R}_2$ by four other variables: ${\bf R}_c$,  the rotational frequency  $\Omega={\bm\nu}\times\dot{\bm\nu}$ and the deviation $r$.

The potential energy is  expanded in a Taylor series to quadratic order around the rest length $a$
\begin{equation}
	\label{eq:pot_taylor}
	U( \left| {\bf R}_1 - {\bf R}_2\right| ) \simeq U(a) + \frac{1}{2} \frac{{\rm d}^2 U}{{\rm d} r^2}\Bigg|_{\left| {\bf R}_1 - {\bf R}_2\right| = a} \!\!\!\!\!\!\!\!\!\!\!\! \!\!\!\!\!\!\!\!\!\!\!\!   r^2 \, ,
\end{equation}
where we have used  $\left| {\bf R}_1 - {\bf R}_2\right|  = a + r$ and $|r| \ll a$. Defining the effective spring constant $k \equiv  \frac{{\rm d}^2 U}{{\rm d} r^2} \left(\left|  {\bf R}_1 - {\bf R}_2\right| = a\right)$, %and moment of inertia $I=Ma^2$, 
the total energy of our dumbbell is written as
%%%
\begin{eqnarray}
		\label{pot-energy}
	H = H_k + U &=& \frac{1}{2} M \vecv_c^2 +\frac{1}{2} I(r) {\Omega}^2 + \frac{1}{2} M \dot{r}^2+ \frac{1}{2} k r^2 \\ \nonumber
	&=& \frac{1}{2} M \vecv_c^2 +\frac{1}{2} I_0 {\Omega}^2 + \frac{1}{2} M \dot{r}^2+ \frac{1}{2} k r^2 + \half Mr\left(2a+r\right)\Omega^2 	\, ,
\end{eqnarray}
%%%%
where $\vecv_c = \dot{\bf R}_c$, $I(r) = M(a+r)^2$, $I_0 = Ma^2$, and the constant contribution $U(a)$ was ignored.
%
%To this, we add the spring constant $k$ to the spring to obtain
%%%%
%\begin{equation}
%	K+U = \frac{1}{2} M V_c^2 +\frac{1}{2} I \dot{\Omega}^2 + \frac{1}{2} M \dot{r}^2+ \frac{1}{2} k r^2 \, .
%\label{pot-energy}
%\end{equation}
%%%%%
%
This equation has four kinetic terms, one potential term, and one term that couples the rotational velocity $\Omega$ with $r$.
Because $r$ has a potential energy coupling, it is not a hydrodynamic variable, but it does couple to $\Omega$ via the moment of inertia $r$ dependence.
In the absence of such coupling, $r$ is an harmonic oscillator with characteristic frequency $\omega = \sqrt{k/M}$. 
(Note that $\omega$ in this section is not related to the vorticity as in the rest of the paper.)
We therefore eliminate $r$ by replacing it with its average value $r_c$, which is obtained by minimizing $H$  with respect to $r$:
\begin{eqnarray}
	\frac{\partial H}{\partial r}\Bigg|_{r=r_c} = M[\omega^2 r + (a+r) \Omega^2 ]=0 \, ,
\end{eqnarray}
such that 
\begin{equation}
 r_c = - a  \frac{\Omega^2}{\omega^2} \frac{1}{1 + \left(\frac{\Omega}{\omega}\right)^2} \simeq   - a  \frac{\Omega^2}{\omega^2} \left( 1 - \frac{\Omega^2}{\omega^2}\right) \, .
\label{r-limit}
\end{equation}
This result implies that $r_c=0$ when $\omega \to \infty$, as it should. 
Note that we have assumed in Eq.~\eqref{eq:pot_taylor} that $|r| \ll a$, such that consistency requires that $\Omega \ll\omega$.
Substituting $r=r_c$ back into the Hamiltonian of Eq.~\eqref{pot-energy} we get
\begin{equation}
	H =  \frac{1}{2} M \vecv_c^2  + \frac{1}{2} I_0\Omega^2 \frac{\omega^2}{\omega^2 + \Omega^2} 
	\simeq   \frac{1}{2} M \vecv_c^2  + \frac{1}{2} I_0\Omega^2 \left( 1 - \frac{\Omega^2}{\omega^2}\right)\, ,
\end{equation}
where the first term is the CM kinetic energy and the second term is the rotational kinetic energy after including the effect of $r$ on the moment of inertia. To leading order in $\Omega/\omega$, the rotational kinetic energy assumes its normal form $I_0  \Omega^2/2$.
%The rotational kinetic energy assumes its normal form $M a^2   \Omega^2/2 $ when $\omega \gg \Omega$.
%The second term of $H$ reduces to zero when $\omega \to 0$ and to $M a^2   \Omega^2/2 $ when $\omega \gg \Omega$.  At fixed $|\omega| $ , $H_r$ reduces to $M a^2   \Omega^2/2 $ in the harmonic small $\Omega$ limit.  
Moreover, we find that the coupling to the non-hydrodynamic variable $r$ does not modify the form of the kinetic energy associated with only the zero-modes (which are the rigid-body translations and rotations).
%harmonic limit of the harmonic $\Omega$ term.  
It does, however, introduce  non-linearities in $\Omega$ which will modify the renormalized harmonic-limit.  The variable $r$ oscillates with frequency $\sqrt{\omega^2+\Omega^2}\simeq\omega\left[1+\left(\Omega^2/\omega^2\right)/2\right]$ about $r_c$  and  thus averages to $r_c$ in the hydrodynamic limit where one is interested in the long-time behavior, i.e., $t \gg 1/\omega$.  When there is dissipation, $r$ will decay to $r_c$ while oscillating.

To exemplify the consequence of driving a finite-frequency mode, it is instructive to add to Eq.~\eqref{pot-energy} a term $-f r$, where $f$ is an external force 
%that can also depend on time. 
that we assume to be constant for simplicity. 
In this case $r_c = \left[ \left(f/M\right)  - a  \Omega^2\right] / \left( \omega^2 + \Omega^2\right)$ and 
\begin{eqnarray}
\nonumber H &=& \frac{1}{2} M \vecv_c^2  + \frac{1}{2} I_0 \frac{\omega^2}{\omega^2+\Omega^2} \left[ 1 + \frac{\alpha^2}{\omega^2}\left(2-\frac{\alpha^2}{\Omega^2}\right) \right] \Omega^2\\
&\simeq&  \frac{1}{2} M \vecv_c^2  + \frac{1}{2} I_0 \left[ \left(1 + \frac{2\alpha^2}{\omega^2}\right) \left(1-\frac{\Omega^2}{\omega^2}\right) - \frac{ \alpha^4 }{ \omega^4} \right] \Omega^2  \, ,
\end{eqnarray}
%
%$H \simeq  \frac{1}{2} M \vecv_c^2  + \frac{1}{2} I_0 \left[ \left(1 + 2f/\left(Ma\omega^2\right)\right) \left(1-\Omega^2/\omega^2\right) - \left( f/\left(Ma\omega^2\right)\right)^2 \right] \Omega^2$, 
where $\alpha^2 \equiv f/(Ma)$ and in the second line we have also dropped a constant term $\sim f^2$.
%where we have dropped a constant term $\sim f^2$. 
We see that the Hamiltonian structure remains intact, and that the external driving of the finite-frequency mode only affects the moment of inertia. Notice that $f$ changes the value of the moment of inertia in the hydrodynamic limit, i.e., it affects the harmonic limit directly.

%In the hydrodynamic limit one is interested in the long-time behavior, i.e., $t \gg 1/\omega$ in which case the effect of the oscillations is averaged to zero.

%This equation has four kinetic terms and one potential term.  The three kinetic  terms  are the same as the kinetic terms in  Eq.~(17) of the main text.  The remaining kinetic term pairs with the static potential energy $k r^2/2$ to produced an harmonic oscillator with characteristic frequency $\omega = \sqrt{k/M}$. In the hydrodynamic limit one is interested in the long-time behavior, i.e., $t \gg 1/\omega$ in which case the effect of the oscillations is averaged to zero.

This phenomena is in a sense universal. We know that frames with sites connected by inextensible rods have $d(d+1)/2$ zero energy modes consisting of $d$ independent rigid translations and $d(d-1)/2$ rigid rotations, independent of what kind of ``bulk'' modes a molecule has~\cite{Mao2018}.  
%Note that there can be internal zero modes in a molecule. These will not average to zero at long times, and could affect the hydrodynamics.
%zero and will contribute to the hydrodynamics similarly to the rigid body translations and rotations.
%
What this example shows, is that in the process of CG, all of the vibrational modes of the molecules (e.g., $r$ in this simple example) are averaged out, and only the molecules' zero-modes {\it may} contribute to the hydrodynamics (finite-frequency modes can only alter molecular parameters). It is therefore important to identify the zero-modes of the molecules before CG.  

A useful characterization of complex molecules is that introduced by James Clerk Maxwell~\cite{Maxwell}, and extended by Calladine~\cite{Calladine},  to describe the strength of what Maxwell called frames consisting of $N$ sites (atoms) connected by $N_B$ stress-supporting bonds (central-force springs). These frames, or complex molecules, generally have $N_0$ zero-energy modes, and $N_S$ states of self-stress.  The latter are structures that can support stress without the exertion of a force on any sites  (e.g., a two-dimensional square with four springs connecting nearest-neighbor sites and two others connecting diagonal sites has one state of self stress).  The Maxwell-Calladine theorem states that $N_0-N_S = dN - N_B$, where $d$ is the system dimensions. For simple molecules $N_s=0$ such that the zero-modes are those associated with rigid body translations and rotations of the molecule, which is what we considered in the main text.
For example, a two-dimensional equilateral triangle has $N=3$ sites, $N_B = 3$, and $N_S=0$ and thus $N_0 = 2 \times 3 - 3 = 3$ corresponding to the two translational and one rotational zero mode that every two-dimensional frame has. Our model diatomic molecules consist of two atoms ($N=2$ ) in $d=3$ connected by a rigid spring ($N_B=1$) with no states of self stress ($N_S=0$).  It, therefore, has $3 N =6$ degrees of freedom, and $N_0 = 6-1=5$, zero modes. This is precisely the number of zero modes for each molecule that appears in $\hat\vecg$ of \eqref{eq:g_s}.  

Note that this discussion  implies that if the fluid molecules have states of self-stress, other ``internal'' zero-modes will have to be considered in the CG process. It is not clear if all zero-modes affect the hydrodynamic equations, or if there are other conditions  on these modes that must be satisfied. We leave this question for future study. In this work we assume no such internal zero-modes exists in the molecules.

\section{Coarse-grained kinetic energy for general complex molecules} \label{sec:general}

Here we extend the model described in the main text to account for a fluid of general complex (rigid) molecules (rather than dumbbells), each of which composed of multiple sub-particles (atoms) with mass $m^{\alpha\mu}$ and momentum $\vecp^{\alpha\mu}$ located at $\vecr^{\alpha\mu}$. As already shown in the supplementary of~\cite{MarLub2021} the coarse-grained hydrodynamic momentum for such fluid is $\vecg = \vecg^c + \half\nabla\times \vecl + \nabla\cdot {\bf A}$, where the definitions of $\vecl$ and ${\bf A}$ are similar to those in the main text of the current paper. The angular momentum density is $\hat{\vecl}(\vecr) = \sum_\alpha \vecl^\alpha \delta(\vecr-\vecr^\alpha)$ with
\begin{eqnarray}
\label{eq:gen3}
\ell^\alpha_i = \varepsilon_{ijk} \sum_{\mu} \Delta p_k^{\alpha\mu}  \Delta r_j^{\alpha\mu} \, ,
\end{eqnarray}
where $\Delta\vecr^{\alpha\mu} \equiv \vecr^{\alpha\mu} - \vecr^\alpha$ and $\Delta\vecp^{\alpha\mu} = m^{\alpha\mu}\partial_t\Delta\vecr^{\alpha\mu}$. The CM momentum $\vecp^\alpha \equiv \sum_\mu \vecp^{\alpha\mu}$ such that $\hat{\vecg}^c(\vecr) \equiv \sum_\alpha \vecp^\alpha \delta(\vecr-\vecr^\alpha)$. We further define $\hat {\bf K}(\vecr) = \sum_\alpha {\bf K}^\alpha \delta\left(\vecr-\vecr^\alpha\right)$, which is a generalization of the alignment tensor and should be thought of as an order parameter (see supplementary of~\cite{MarLub2021}) with ${\bf K}^\alpha = \sum_\mu m^{\alpha\mu} \Delta\vecr^{\alpha\mu} \Delta\vecr^{\alpha\mu} = \frac{3}{2} \left( {\bf R}^\alpha +\frac{1}{3}{\bf I}^\alpha \right)$ and~\cite{Lubensky2003}
\begin{eqnarray}
\label{eq:gen1}
& &R_{ij}^\alpha = \sum_{\mu} m^{\alpha\mu} \left[ \Delta r_i^{\alpha\mu} \Delta r_j^{\alpha\mu} - \frac{1}{3}\left(\Delta \vecr^{\alpha\mu}\right)^2\delta_{ij} \right] \, , \\
\label{eq:gen1a}
& &I_{ij}^\alpha = \sum_{\mu} m^{\alpha\mu} \left[ \left(\Delta \vecr^{\alpha\mu}\right)^2\delta_{ij}  - \Delta r_i^{\alpha\mu} \Delta r_j^{\alpha\mu} \right] \, .
\end{eqnarray}
Then, $\hat{{\bf A}}(\vecr) = \half\sum_\alpha \dot{\bf K}^\alpha \delta(\vecr-\vecr^\alpha)$. 

Notably, the fundamental PBs (see Eq.~(25) in Materials and Methods) are not modified, hence, the only thing left to find is the form of the Hamiltonian, which we will show to have the same form as in Eq.~(\ref{m7}) such that the dynamics we use in the main text applies also for a fluid composed of general complex molecules. We begin with the Hamiltonian of Eq.~\eqref{m1} where now the kinetic energy density ${\cal H}_k$, defined as before via $H_k = \int \D\vecr {\cal H}_k$ is:
\begin{eqnarray}
\label{eq:gen2}
{\cal H}_k(\vecr) =  \sum_{\alpha\mu} \frac{1}{2m^{\alpha\mu}} \left( \frac{m^{\alpha\mu}}{M^\alpha}\vecp^\alpha + \Delta\vecp^{\alpha\mu}\right)^2 
\delta\left(\vecr-\vecr^\alpha-\Delta\vecr^{\alpha\mu}\right) \, .
\end{eqnarray}
Here $\vecp^{\alpha\mu} = \vecp^\alpha\left(m^{\alpha\mu}/M^\alpha\right) + \Delta\vecp^{\alpha\mu}$ and $M^\alpha = \sum_\mu m^{\alpha\mu}$ is the molecule mass. Assuming the molecules are small compared to typical distances (long wavelength limit), $\left|\Delta\vecr^{\alpha\mu}\right|\ll \left|\vecr-\vecr^\alpha\right|$ we write $\delta\left(\vecr-\vecr^\alpha-\Delta\vecr^{\alpha\mu}\right) \simeq \delta\left(\vecr-\vecr^\alpha\right) - \Delta\vecr^{\alpha\mu} \cdot \nabla \delta\left(\vecr-\vecr^\alpha\right)$ such that Eq.~\eqref{eq:gen2} becomes
\begin{eqnarray}
\label{eq:gen2a}
{\cal H}_k(\vecr) =  \sum_\alpha \frac{\left(\vecp^\alpha\right)^2}{2M^{\alpha}} \delta\left(\vecr-\vecr^\alpha\right) 
+ \sum_{\alpha\mu} \left[ \frac{\left(\Delta\vecp^{\alpha\mu}\right)^2}{2m^{\alpha\mu}} - \frac{\vecp^\alpha\cdot \Delta\vecp^{\alpha\mu}}{M^{\alpha}} \Delta\vecr^{\alpha\mu} \cdot \nabla \right] \delta\left(\vecr-\vecr^\alpha\right)\, ,
\end{eqnarray}
where we have used $\sum_\mu m^{\alpha\mu}  \Delta\vecr^{\alpha\mu} = \sum_\mu\Delta\vecp^{\alpha\mu} = 0$ and neglected a term $\sim \left(\Delta\vecp^{\alpha\mu}\right)^2 \Delta\vecr^{\alpha\mu}$ that is proportional to the cube of the molecule size (we only keep quadratic order). The second term in Eq.~\eqref{eq:gen2a} can be written in more familiar terms: 
\begin{eqnarray}
\label{eq:gen2b}
\sum_{\mu}  \frac{\left(\Delta\vecp^{\alpha\mu}\right)^2}{2m^{\alpha\mu}}  = \half \ell_i^\alpha \left(I_{ij}^\alpha\right)^{-1} \ell_j^\alpha\, ,
\end{eqnarray}
with ${\bm \Omega}^\alpha =  \left(I_{ij}^\alpha\right)^{-1} \ell_j^\alpha$ being  the angular velocity of a molecule around its CM which for rigid molecules obeys $\partial_t \Delta\vecr^{\alpha\mu} = -\Delta\vecr^{\alpha\mu}\times {\bm \Omega}^\alpha$. The third term of Eq.~\eqref{eq:gen2a} can also be simplified with the help of Eqs.~\eqref{eq:gen3}-\eqref{eq:gen1a}:
\begin{eqnarray}
\label{eq:gen2c}
\sum_{\mu} \frac{\vecp^\alpha\cdot \Delta\vecp^{\alpha\mu}}{M^{\alpha}} \Delta\vecr^{\alpha\mu} =  \frac{p_i^\alpha}{2M^\alpha}  \left(\dot{K}_{ij}^\alpha -  \varepsilon_{ijk}\ell_k^\alpha \right) \, .
\end{eqnarray}

Substituting Eqs.~\eqref{eq:gen2b} and \eqref{eq:gen2c} into \eqref{eq:gen2a} we get
\begin{eqnarray}
\label{eq:gen4}
{\cal H}_k(\vecr) =  \sum_\alpha \Bigg[ \frac{\left(\vecp^\alpha\right)^2}{2M^{\alpha}} 
+ \half \ell_i^\alpha \left(I_{ij}^\alpha\right)^{-1} \ell_j^\alpha 
+ \frac{1}{2M^\alpha} \nabla \cdot\left(\vecl^\alpha \times \vecp^\alpha - \vecp^\alpha\cdot \dot{\bf K}^\alpha \right) \Bigg] \delta\left(\vecr-\vecr^\alpha\right)\, .
\end{eqnarray}
The last step is to coarse-grain using the method described in Materials and Methods, which gives a very similar Hamiltonian to the one of Eq.~(\ref{m7}),
\begin{eqnarray}
\label{eq:gen5}
 H_k = \int\D\vecr \Bigg[ \frac{\left(\vecg^c\right)^2}{2\rho} + \half \vecl \cdot {\bf I}^{-1} \cdot \vecl 
+ \half\nabla \cdot \left(\vecl \times \vecv -\vecv\cdot{\bf A}\right) \Bigg]  \, ,
\end{eqnarray}
where ${\bf A}$ here is a generalization of the one in the main text (for diatomic molecules ${\bf K}^\alpha = {\bf Q}^\alpha$). 
%In the case of diatomic molecules this Hamiltonian coincides with the Hamiltonian used in the main text. 

\section{Hydrodynamic momentum density dynamics} \label{sec:total}

Deriving the dynamics of the hydrodynamic momentum density (Eq.~(5) of the main text) from that of the CM momentum density (Eqs.~(3)-(4) of the main text) requires some non-trivial manipulation that we details in this section.

Starting from Eqs.~(3)-(4) of the main text and using Eq.~(2) of the main text we can write the dynamics of the hydrodynamic momentum:
\begin{eqnarray}
\label{pb2}
\nonumber\dot{g}_i &=& \dot{g}^c_i + \half\varepsilon_{ijk} \nabla_j \dot{\ell}_k 
 -\nabla_j \left(v_j g_i\right) + \nabla_j \left(  v_j g_i - v_j^c g_i^c\right) -\nabla_i P 
+ \eta^e_{ijkl}\nabla_lv_k^c \\
& \qquad & -\half\varepsilon_{ijn} \nabla_j \nabla_m \left(v_m^c \ell_n\right) +\half\varepsilon_{ijk} \nabla_j \tau_k + f_i\, .
\end{eqnarray}
Converting $\nabla_j(g^c_i v^c_g)$ to $\nabla_j(g_i v_g)$ with the help of Eq.~(2) of the main text we  arrive at
\begin{eqnarray}
\label{pb5}
\dot{g}_i + \nabla_j(g_i v_g) = f_i -\nabla_i P
+ \nabla_j \left( \left[ \eta^e_{ijkl} - \half \ell_n \left( \varepsilon_{iln}\delta_{jk} + \varepsilon_{jln}\delta_{ik}\right) \right]   \nabla_l v_k  +\half\varepsilon_{ijk} \tau_k  \right) \, ,
\end{eqnarray}
where we have dropped a non hydrodynamic term $\sim \nabla \left(\nabla\times\vecl\right)^2$, and we have also took advantage of the identity $\nabla_j \varepsilon_{jln} \nabla_l (\ell_n v_i) =  \nabla_j\left[v_i \left(\nabla\times\vecl\right)_j + \varepsilon_{jln} \ell_n \nabla_l  v_i\right] = 0$.
The final step is to separate the velocity gradient tensor to its symmetric and antisymmetric parts. Then, the symmetric part of the first term in the square brackets of Eq.~\eqref{pb5} gives the odd viscosity $-(\ell_n/4)\gamma^o_{ijkl;n}$, while its antisymmetric part can be written as
\begin{eqnarray}
\label{eq:antisymm1}
-\frac{\ell_n}{4}\left[ \varepsilon_{iln} \delta_{jk} +\varepsilon_{jln} \delta_{ik} - \varepsilon_{ikn} \delta_{jl} - \varepsilon_{jkn} \delta_{il} \right] \nabla_l v_k 
=-\frac{\ell_n}{4}\varepsilon_{lkm} \varepsilon_{mst} \left(\varepsilon_{iln} \delta_{jk} + \varepsilon_{jln} \delta_{ik} \right) \nabla_s v_t \, .
\end{eqnarray}
We then use
\begin{eqnarray}
\label{eq:antisymm2}
\varepsilon_{lkm} \varepsilon_{mst} \varepsilon_{iln} \delta_{jk} \nabla_s v_t = \left(\varepsilon_{ilk} \delta_{nj} - \varepsilon_{lkn} \delta_{ij} \right) \nabla_l v_k \, ,
\end{eqnarray}
to obtain
\begin{eqnarray}
\label{eq:antisymm3}
-\frac{\ell_n}{4}\left[ \varepsilon_{iln} \delta_{jk} +\varepsilon_{jln} \delta_{ik} - \varepsilon_{ikn} \delta_{jl} - \varepsilon_{jkn} \delta_{il} \right] \nabla_l v_k 
= -\frac{\ell_n}{4}\left(\varepsilon_{ilk}\delta_{nj}  + \varepsilon_{jlk}\delta_{ni}  - 2\varepsilon_{lkn}\delta_{ij} \right) \nabla_l v_k \, ,
\end{eqnarray}
where we also utilized Eq.~\eqref{eq:antisymm2} with interchanging $i\leftrightarrow j$. This  gives Eq.~(5) of the main text.

\section{Elimination of angular momentum in the CM dynamics} \label{sec:elimination}
In this section we eliminate the spin angular momentum (SAM) using Eq.~(3) of the main text and write the CM dynamics of Eq.~(4) of the main text after elimination of $\vecl$. We then obtain Eq.~(5) of the main text for the hydrodynamic momentum dynamics with $\vecl\to\vecl_0$. This procedure is equivalent to what we do in the main text, namely, writing the hydrodynamic momentum dynamics and then eliminate $\vecl$.

We start by solving for $\Omega$ in the hydrodynamic limit from Eq.~(3) of the main text:
	\begin{equation}
		\label{eq:ell_dyn}
		\dot{\ell}_i(\vecr) + \nabla_j\left(\ell_i v^c_j\right) = - \Gamma \left(\Omega_i - \omega^c_i\right) + \tau_i  \, ,
	\end{equation}
	where $\vecl = \rho\tilde{I} {\bm\Omega}$. We  write ${\bm\Omega} = {\bm\Omega}_0 + \delta{\bm\Omega}$ and $\rho = \rho_0 + \delta\rho$ while assuming that $\delta{\bm \Omega}$ and $\delta\rho$ are of the order of ${\bm\nabla}{\bf v}^c$. Here we include also possible friction with the surface such that ${\bm\tau} = {\bm{\tilde \tau}} -\Gamma^{\rm ex} {\bm\Omega}$. With the help of the continuity equation we then have:
	\begin{eqnarray}
		\label{eq:Omega_0}
		& &\left(\partial_t + \Gamma^T / I_0 \right) {\bm\Omega}_0 = \tilde{\bm\tau}(\vecr,t)/I_0 \, , \\
		\label{eq:delta_Omega}
		& &\left(\partial_t + \Gamma^T / I_0 \right) \delta{\bm\Omega} = {\bm\Phi}(\vecr,t)/I_0 \, ,
	\end{eqnarray}
	where $ {\bm\Phi}(\vecr,t) \equiv \Gamma{\bm\omega}^c - I_0 \vecv^c\cdot\nabla{\bm\Omega}_0$, $I_0 =  \tilde{I}\rho_0$, and $\Gamma^T = \Gamma + \Gamma^{\rm ex}$. The solution to Eqs.~\eqref{eq:Omega_0}-\eqref{eq:delta_Omega} is
	\begin{eqnarray}
		\label{eq:Omega_0_sol}
		& &{\bm\Omega}_0(\vecr,t) = {\bm\Omega}_0(t=0) \e^{-T} \!\! + \!\! \int_{0}^T \frac{\tilde{\bm\tau}(\vecr,u)}{\Gamma^T} \e^{-(T-u)} \D u  , \quad\\
		\label{eq:delta_Omega_sol}
		& &\delta{\bm\Omega}(\vecr,t) = \delta{\bm\Omega}(t=0) \e^{-T} \!\!+ \!\!\int_{0}^T \!\! \frac{{\bm\Phi}(\vecr,u) }{\Gamma^T} \e^{-(T-u)} \D u   , \quad
	\end{eqnarray}
	with $T\equiv \Gamma^T t / I_0$. In the hydrodynamic limit $T\gg 1$ (which is equivalent to the zero-frequency limit) we are left with 
	\begin{eqnarray}
		\label{eq:Omega_sol_final}
		{\bm\Omega}_0(\vecr,t) \simeq\tilde{\bm\tau}(\vecr,t) / \Gamma^T  \,\,\, ; \,\,\, \delta{\bm\Omega}(\vecr,t) \simeq {\bm\Phi}(\vecr,t) / \Gamma^T   \, ,
	\end{eqnarray}
	such that $\vecl^0 = I_0 \tilde{\bm\tau}(\vecr,t) / \Gamma^T$. This solution is valid for any $\tilde{\bm \tau}$, even if it is inhomogeneous and not constant in time.
	It essentially sets $\dot{\Omega}=0$ such that with the aid of the continuity equation we find that
	\begin{eqnarray}
		\label{eq:dot_ell}
		\dot{\bm\ell} \simeq -\tilde{I}{\bm\Omega}_0 \nabla\cdot(\rho_0\vecv^c)   \, .
	\end{eqnarray}
This result suggests that a description in terms of the CM momentum alone will not be sufficient as it will not obey conservation of total angular momentum.  This is because the SAM within a fluid parcel is not constant, $\dot{\ell}_i + \nabla_j\left(\ell_i v^c_j\right) = I_0 \vecv^c \cdot \nabla {\bm \Omega}_0$. 
Only when ${\bm \Omega}_0$ is homogeneous is the SAM constant within a fluid parcel and the CM momentum description obeys the balance of angular momentum. We will see this explicitly below.

	\subsection{CM stress tensor} \label{sec:CM_stress}
	Let us now write the CM dynamics after elimination of $\vecl$. Before elimination of $\vecl$ the CM dynamics follows Eq.~(4) of the main text 
	$\dot{g}_i^c + \nabla_j \left(v_j^c g_i^c\right) = \nabla_j\sigma^c_{ij} + f_i$ where the CM stress tensor is:
	\begin{equation}
		\label{eq:stress1}
		\sigma^c_{ij} =   - P \delta_{ij} + \eta^e_{ijkl} \nabla_l v^c_k + \frac{\Gamma}{2}\varepsilon_{ijk} \left(\Omega_k - \omega_k^c\right)  \, .
	\end{equation}
	It is convenient to use Eq.~\eqref{eq:ell_dyn} to simplify the antisymmetric part of the stress:
	\begin{equation}
		\label{eq:stress2}
		\sigma^c_{ij} =   - P \delta_{ij} + \eta^e_{ijkl} \nabla_l v^c_k + \half \varepsilon_{ijn} \left[( \tau_n - \dot{\ell}_n - \nabla_k\left(\ell_n v^c_k\right) \right]  \, .
	\end{equation}
	Then, by using Eq.~\eqref{eq:dot_ell} we find that 
	\begin{equation}
		\label{eq:stress_CM}
		\sigma^c_{ij} =   - P \delta_{ij} + \eta^e_{ijkl} \nabla_l v^c_k + \half \varepsilon_{ijn} \left( \tau_n - I_0 \vecv^c\cdot\nabla \Omega_n^0 \right)   \, .
	\end{equation}

	It is clear that even after elimination of $\vecl$, odd viscosity does not appear in the CM stress tensor. Only when the active torques are inhomogeneous, the angular velocity affects the CM stress by adding an antisymmetric term to it. This term, not only seems to break Galilean invariance in the CM dynamics, but also breaks conservation of angular momentum (the only antisymmetric term allowed in the stress tensor is the external torque, $\half \varepsilon_{ijn}  \tau_n$). The latter explicitly shows that some angular momentum is ``missing'' in the CM description, which is also consistent with the fact that even after relaxation of $\vecl$ the SAM of a fluid parcel is changing with time, see Eq.~\eqref{eq:dot_ell}.

	In the case of homogeneous $\tilde{\bm\tau}$ the CM stress tensor can be written in the regular form~\cite{LubenskyBook} with the addition of an antisymmetric stress directly due to the external torque:  
\begin{equation}
	\label{eq:stress_CM_const}
	\sigma^c_{ij} =   - P \delta_{ij} + \eta^e_{ijkl} \nabla_l v^c_k + \half \varepsilon_{ijn}  \tau_n   \, .
\end{equation}
As discussed in the previous subsection, only in this case does the CM momentum obeys the balance of angular momentum. 
In the case that there is also no external friction $\Gamma^{\rm ex}=0$, the antisymmetric term vanishes in the bulk, such that in the bulk the CM stress is symmetric and therefore conserve angular momentum.
Remarkably, as we shall see in the next subsection, even in this simplified case, the flux of hydrodynamic  momentum (which is the hydrodynamic momentum stress tensor) is different from the flux of CM momentum (the CM stress tensor above). This is manifested in the odd terms that only emerge in the hydrodynamic momentum dynamics.

	\subsection{Hydrodynamic momentum stress tensor} \label{sec:total_stress}
	Following the previous subsection, we show here how to obtain Eq.~(5) of the main text after elimination of $\vecl$ in the CM dynamics. The derivation is similar to that in Sec.~\ref{sec:total} where the only difference is the change of  ${\bm \sigma}^c$ (see Eq.~\eqref{eq:stress_CM}) due to elimination of $\vecl$.  We start from writing a modified Eq.~\eqref{pb2}:
	\begin{eqnarray}
		\label{eq:total1}
		\dot{g}_i + \nabla_j \left(v_j g_i\right) = \dot{g}^c_i + \half\varepsilon_{ijk} \nabla_j \dot{\ell}_k + \nabla_j \left(v_j g_i\right) 
		= \nabla_j \left( v_j g_i - v_j^c g_i^c\right)  + \half\varepsilon_{ijk} \nabla_j \dot{\ell}_k  + \nabla_j\sigma^c_{ij} +f_i  \, .
	\end{eqnarray}
	Then, following the same derivation of Eq.~\eqref{pb5} and using Eq.~\eqref{eq:stress_CM} we have:
	\begin{eqnarray}
		\label{eq:total2}
		\nonumber\dot{g}_i + \nabla_j(g_i v_g) &=& f_i -\nabla_i P + \half\varepsilon_{ijk} \nabla_j \dot{\ell}_k \\
		&+& \nabla_j \left[ \left( \eta^e_{ijkl} - \half \ell_n^0 \left[ \varepsilon_{iln}\delta_{jk} + \varepsilon_{jln}\delta_{ik}\right] \right)   \nabla_l v_k   
		+ \half\varepsilon_{ijn} \left( \tau_n +\tilde{I} \Omega_n^0 \nabla_k\left(\rho_0 v_k\right)  \right) \right] \, . 
	\end{eqnarray}
	Substituting Eq.~\eqref{eq:dot_ell} we finally get
	\begin{eqnarray}
		\label{eq:total3}
		\dot{g}_i + \nabla_j(g_i v_g) = f_i -\nabla_i P 
		+ \nabla_j \left[ \left( \eta^e_{ijkl} - \half \ell_n^0 \left[ \varepsilon_{iln}\delta_{jk} + \varepsilon_{jln}\delta_{ik}\right] \right)   \nabla_l v_k   
		+ \half\varepsilon_{ijn}  \tau_n \right]  \,,
	\end{eqnarray}
	which is the same as Eq.~\eqref{pb5} with $\vecl\to\vecl^0=I_0\tilde{\bm\tau}/\Gamma^T$. The derivation of Eq.~(5) of the main text from this point is the same as after Eq.~\eqref{pb5}.

\section{Simple examples for our coarse-graining process} \label{sec:cg}

To verify the validity of our coarse-graining approach let us apply it to the kinetic energy density of a monoatomic gas
\begin{eqnarray}
\label{e4a}
{\cal H}_k(\vecr) = \sum_\alpha \frac{\left(\vecp^\alpha\right)^2}{2m} \delta(\vecr-\vecr^\alpha) \, .
\end{eqnarray}
%
%${\cal H}_k(\vecr) = \sum_\alpha \left(\vecp^\alpha\right)^2/(2m) \delta(\vecr-\vecr^\alpha)$. 
This is done straightforwardly by writing $\boldsymbol{A}^\alpha = \vecp^\alpha$ and $\boldsymbol{B}^\alpha = \vecp^\alpha/(2m)$. Then, using Eq.~(22) in Materials and Methods the coarse-grained kinetic energy is
\begin{eqnarray}
\label{e4}
{\epsilon}(\vecr) =   \frac{\vecg(\vecr)^2}{2\rho(\vecr)} \, ,
\end{eqnarray}
as expected. Here  $\rho(\vecr)=m n(\vecr)$ and $\vecg(\vecr) = \sum_\alpha \vecp^\alpha W(\vecr-\vecr^\alpha)$. 

Note that we can apply our prescription also for functions incorporating more microscopic variables. For example, if the particles' masses are not equal,
\begin{eqnarray}
\label{e5a}
{\cal H}_k(\vecr) = \sum_\alpha \frac{\left(\vecp^\alpha\right)^2}{2m^\alpha} \delta(\vecr-\vecr^\alpha) \, .
\end{eqnarray}
%
%${\cal H}_k(\vecr) = \sum_\alpha \left(\vecp^\alpha\right)^2/(2m^\alpha) \delta(\vecr-\vecr^\alpha)$. 
Let us write in general 
\begin{eqnarray}
\label{e5b}
\hat{\cal O}(\vecr) = \sum_\alpha{\bf A}^\alpha \left({\bf C}^\alpha\right)^{-1} {\bf B}^\alpha \delta(\vecr-\vecr^\alpha) \, .
\end{eqnarray}
%
%$\hat{\cal O}(\vecr) = \sum_\alpha{\bf A}^\alpha \left({\bf C}^\alpha\right)^{-1} {\bf B}^\alpha \delta(\vecr-\vecr^\alpha)$. 
Then, after using $\bar{\bf B}(\vecr) = {\bf B}(\vecr)/n(\vecr)$ as we do in Materials and Methods, and replacing $\left({\bf C}^\alpha\right)^{-1} $ with $\overline{\left({\bf C}^\alpha\right)^{-1}}$ within the CG volume:
\begin{eqnarray}
\label{e5b1}
\overline{\left({\bf C}^\alpha\right)^{-1}} \equiv \frac{ \int_{\Delta V} \D{\bf u} \sum_\alpha \left({\bf C}^\alpha\right)^{-1}  W(\vecr- {\bf u} - \vecr^\alpha)}
{\int_{\Delta V} \D{\bf u}  \sum_\alpha  W(\vecr- {\bf u} - \vecr^\alpha)} 
\simeq  \left[\bar{\bf C}(\vecr)\right]^{-1} \, ,
\end{eqnarray}
we find that 
\begin{eqnarray}
\label{e5c}
{\cal O}(\vecr) \simeq {\bf A}(\vecr) \left[{\bf C}(\vecr)\right]^{-1} {\bf B}(\vecr) \, .
\end{eqnarray}
%
%${\cal O}(\vecr) \simeq {\bf A}(\vecr) \left[{\bf C}(\vecr)\right]^{-1} {\bf B}(\vecr)$. 
For the example of the kinetic energy above this clearly gives ${\cal H}_k \simeq \vecg^2/(2\rho)$ as desired.
Note that in Eq.~\eqref{e5b1}  we have used
\begin{eqnarray}
\label{e5b2}
\frac{1}{\Delta V} \int_{\Delta V} \D{\bf u} \sum_\alpha \left({\bf C}^\alpha\right)^{-1}  W(\vecr- {\bf u} - \vecr^\alpha) \simeq n(\vecr) \left[\bar{\bf C}(\vecr)\right]^{-1} \, .
\end{eqnarray}

Next we want to examine coarse-graining of functions that include gradients such as
\begin{eqnarray}
\label{e5}
\hat{\boldmath{\cal L}}(\vecr) = \sum_\alpha \boldsymbol{A}^\alpha \boldsymbol{B}^\alpha  \nabla \delta(\vecr-\vecr^\alpha) \, . 
\end{eqnarray}
Clearly the integrated value of $\hat{\cal L}(\vecr) $ must give a boundary term because:
\begin{eqnarray}
\label{e6}
\boldsymbol{L} \equiv \int\D\vecr \hat{\boldmath{\cal L}}(\vecr) = \int \D\vecr \nabla \left[ \sum_\alpha \boldsymbol{A}^\alpha \boldsymbol{B}^\alpha   \delta(\vecr-\vecr^\alpha) \right]  
=\oint_S \D\boldsymbol{S} \sum_\alpha \boldsymbol{A}^\alpha \boldsymbol{B}^\alpha   \delta(\boldsymbol{s}-\boldsymbol{s}^\alpha)\, .
\end{eqnarray}
Within our coarse-graining scheme we have
\begin{eqnarray}
\label{e7}
\hat{\boldmath{\cal L}}(\vecr)  \simeq \nabla \left[\frac{\boldsymbol{A}(\vecr)\boldsymbol{B}(\vecr)}{n(\vecr) }\right]\, ,
\end{eqnarray}
which gives a boundary term as it should.

%\section{Kinetic theory vs Irving-Kirkwood} \label{sec:EP}

\section{Exceptional points} \label{sec:EP}

Our analysis showed that the non-Hermitian matrices characterizing the excitations of active systems of rotating objects exhibit an exceptional point at which two eigenvalues coalesce and the system appears to ``lose'' an eigenvalue. In terms of the differential equation we are solving this means we lost a fundamental solution, which is of course not true.
In much of the literature (see e.g., Refs.~\cite{Vitelli2021,Vitelli_review_topological, Heiss2012}) the discussion on exceptional points (EPs) was focused on the EPs marking a breaking of the (generalized) ${\cal PT}$ symmetry and non-reciprocal phase transition~\cite{Vitelli2021}. 
Here follow this recent work that explains EPs through the textbook example of a damped harmonic oscillator (Ref.~\cite{fernandez2018} is the most elaborated, but see also the SI of Ref.~\cite{Vitelli2021} and Ref.~\cite{Vitelli_review_topological}) and explain where is the ``lost'' fundamental solution.

\subsection{Exceptional points in a damped harmonic oscillator} \label{EP_harmonic} 

Let us start with the equation of a one-dimensional damped harmonic oscillator:
%%%%%
\begin{equation}
\ddot{x} + \gamma \dot{x}  +  \omega_0^2 x = 0 \, .
\label{eq:EOM}
\end{equation}
Here, $\omega_0$ is the natural frequency of the undamped oscillator and $\gamma = \Gamma/m$, the inverse decay time, is the friction coefficient $\Gamma$ divided by particle mass $m$.  Following standard procedure, we assume a solution of the form $x(t) \propto e^{-\alpha t}$ which leads to 
%%%
\begin{equation}
\alpha^2  + \gamma \alpha +\omega_0^2 = 0
\label{eq:alpha}
\end{equation}
%%%%
with two solutions for $\alpha$,
%%%%%
\begin{equation}
\alpha_{\pm} = \frac{1}{2}\left( \gamma \pm \sqrt{ \gamma^2 - 4 \omega_0^2 } \right) ,
\label{eq:alpha-2}
\end{equation}
%%%%
and associated eigenfunctions, $\exp (-\alpha_{\pm} t)$.
When $\gamma^2 >  4 \omega_0^2$, the over-damped regime,  the two modes decay exponentially, and when $\gamma^2 <  4 \omega_0^2$,  both modes oscillate and decay (underdamped regime).  When $\gamma^2 = 4 \omega_0^2$, there is only one solution, $\alpha = \gamma/2$, and the system is at the exceptional point. Of course even when $\gamma^2 = 4 \omega_0^2$, Eq.~(\ref{eq:EOM}) must have two solutions, the first of which is the simple exponential one just discussed and the second of which can be found by considering the limit as the EP is approached.  To this end, we set
%%%%%
\begin{equation}
\alpha_{\pm} = \frac{1}{2} \gamma \pm  \sqrt{\Delta} ,
\label{eq:Delta}
\end{equation}
%%%%
where $\Delta = (\gamma^2/4) -  \omega_0^2$, and we construct two fundamental solutions using the sum and difference of the two eigenfunctions:
%%%%
\begin{align}
& \psi_1(t)  = \frac{1}{2}  \left(e^{-\gamma t/2 +\sqrt{\Delta} t}  +  e^{-\gamma  t/2-\sqrt{\Delta} t} \right)\, , \\
& \psi_2(t)  = \frac{1}{2\sqrt{\Delta}} \left(e^{-\gamma t/2 +\sqrt{\Delta} t}  -  e^{-\gamma  t/2-\sqrt{\Delta} t} \right)\, .
\end{align}
The first fundamental solution, $\psi_1$  has a well-defined $\Delta \rightarrow 0$ limit that is $e^{-\gamma t/2}$. The second fundamental solution, $\psi_2$, is divided by $\sqrt{\Delta}$ to give a proper limit as $\Delta \rightarrow 0$, which is $t e^{-\gamma t/2}$.  Thus we have two linearly independent solutions at the EP:
%%%%%%
\begin{equation}
\psi_1(t)= e^{-\gamma t/2}, \qquad \text{and} \qquad \psi_2(t) = t e^{-\gamma t/2} .
\end{equation}
%%%%%
We can then write a general solution to Eq.~\eqref{eq:EOM} (at the EP) for some initial condition $x(t=0)=x_0$ as $x(t) = a \psi_1(t) + b \psi_2(t)$, where $a$ and $b$ are determined from the initial condition.

Though linearly independent, these $\psi_1$ and $\psi_2$ are not orthogonal because $\int_0^\infty dt \psi_1(t) \psi_2(t) = \gamma^{-2} $. Although not necessary, one can create an orthogonal pair by adding to $\psi_2(t)$ a term proportional to $\psi_1$.  A trivial exercise establishes that the function $\tilde{\psi}_2 = (1-\gamma t) e^{-\gamma t /2} = \psi_1 - \gamma \psi_2$ is orthogonal to $\psi_1$.
%
%%\begin{equation}
%%\int_0^\infty dt \psi_1(t) \psi_2(t) = \int_0^\infty dt t e^{-\gamma t} = \gamma^{-2}  
%%\end{equation}
%%is not zero.  
%
%%%%%%
%\begin{equation}
%\tilde{\psi}_2 = (1-\gamma t) e^{-\gamma t /2} = \psi_1 - \gamma \psi_2 
%\end{equation}
%%%%%%
%is orthogonal to $\psi_1$.
%%

\subsection{Exceptional points using matrix representation} \label{EP_mat}

The damped oscillator equation can be couched in terms of a matrix operating on the  vector $\uv(t)= \left[x(t), v(t)= \dot{x}(t)\right]^T$, which may be more familiar to those who study exceptional points.  The equation of motion for $\uv (t)$ is
%%%%%
\begin{equation}
\label{eq:eom_matrix}
\dot{\uv} = \left(\begin{array}{cc}
0 & 1   \\
-\omega_0^2  & - \gamma 
\end{array}  \right)  \uv
\equiv \Mv \,\, \uv  \,\,.
\end{equation}
$\Mv$ has two eigenvectors, $\uv^\pm(t)= \left[1, -\alpha_\pm\right]^T$, associated with the two eigenvalues $\alpha_{\pm}$. 
%%%%%
%\begin{equation}
%\wv^{\pm} = \left( 
%\begin{array}{c}
%1 \\
%-\alpha_{\pm}
%\end{array}
%\right) \, 
%\end{equation}
%%%%%
%associated with the two eigenvalues $\alpha_{\pm}$. 
%
At the EP, where $\Delta=0$, the matrix $\Mv$ is defective (not diagonalizable) and has only one eigenvector $\uv_1 = \left[1, -\gamma/2\right]^T$. It does not have a set of linearly independent eigenvectors that form a basis.

Nevertheless, in general, for any $n \times n$ matrix one can form a basis using {\it generalized eigenvectors} generated by Jordan chains. For a generic matrix $\Mv$ and eigenvalue $\lambda$ a generalized eigenvector of rank $m$ is defined as
\begin{eqnarray}
\left[\Mv - \lambda \bvec{I}\right]^m\bvec{v}_m = 0 \quad ; \quad 
\left[\Mv - \lambda \bvec{I}\right]^{m-1}\bvec{v}_m \neq 0 \, ,
\end{eqnarray}
where $\vecv_1$ (a generalized eigenvector of rank 1) is clearly an ordinary eigenvector. Once one knows a generalized eigenvector $\vecv_m$ the Jordan chain of generalized eigenvectors can be found using $\left[\Mv - \lambda \bvec{I}\right]\bvec{v}_{m+1} = \bvec{v}_{m}$. The generalized eigenvectors can be used as a generalized modal matrix $\bvec{P}$ (a matrix in which the columns are the generalized eigenvectors) to find its Jordan normal form~\cite{fernandez2018} using $\bvec{J} = \bvec{P}^{-1} \Mv \bvec{P}$. The Jordan normal form is a block diagonal matrix, where each block is a Jordan block -- it has the eigenvalues on the diagonal, ones on the superdiagonal and zeros elsewhere. This form simplifies matrix functions (similarly to matrix functions on diagonal matrices).

In our example we solve $\left[\Mv - \lambda \bvec{I}\right]\bvec{u}_2 = \bvec{u}_1$ for $\lambda = \gamma/2$ and find the generalized eigenvector $\uv_2 = \left[0,1\right]^T$.
With the help of the modal matrix 
\begin{equation}
\bvec{P} = \left(\begin{array}{cc}
1 & 0   \\
-\gamma/2  & 1 
\end{array}  \right)  \, ,
\end{equation}
we can write $\Mv$ in Jordan normal form
\begin{equation}
\bvec{J} = \bvec{P}^{-1} \Mv \bvec{P} = \left(\begin{array}{cc}
-\gamma/2 & 1   \\
0  & -\gamma/2 
\end{array}  \right)  \, .
\end{equation}
This matrix is a Jordan block of size two, hence, the EP is of order two (in general a Jordan block of size $m$ is related to an EP of order $m$~\cite{Vitelli2021}).

Notice that the generalized eigenvectors we have found $\uv_1$ and $\uv_2$ are not orthogonal. It is, however, straightforward to find an orthogonal generalized eigenvector because $\tilde\uv_2 = \uv_2 + c \uv_1$  ($c$ being a scalar) is also a generalized eigenvector, and for $c = -\uv_1\cdot\uv_2 / \uv_1^2$ we get that $\uv_1\cdot\tilde\uv_2 = 0$. 
Now that we know the generalized eigenvectors, we can write the formal solution to Eq.~\eqref{eq:eom_matrix}:
\begin{eqnarray}
\uv(t) = \exp\left(\Mv t\right) \uv_0 \, ,
\end{eqnarray}
where $\uv_0 = \uv(t=0)$, and since $\uv_1$ and $\tilde\uv_2$ span the space we can write $\uv_0 = a\uv_1 + b\tilde\uv_2$. Next, we wish to be able to write the general solution to Eq.~\eqref{eq:eom_matrix} as in Sec.~\ref{EP_harmonic}. % $\uv(t) = af_1(t)\uv_1 + b{f}_2(t)\tilde\uv_2$. 
To do so we calculate
\begin{eqnarray}
\nonumber& &\exp\left(\Mv t\right) \uv_1 = \sum_{n=0}^\infty \frac{\Mv^n}{n!} t^n \uv_1 =  \sum_{n=0}^\infty \frac{\lambda^n}{n!} t^n \uv_1 = e^{\lambda t}\uv_1\, ,\\
& &\exp\left(\Mv t\right) \uv_2 =  \uv_2 + \sum_{n=1}^\infty \frac{\Mv^{n-1}}{n!} t^n \left(\uv_1 + \lambda\uv_2\right) 
= e^{\lambda t}\uv_2 + \sum_{n=1}^\infty  \frac{\lambda^{n-1}}{n!}  t^n n \uv_1 = e^{\lambda t} \left( \uv_2 + t\uv_1\right)\, ,
\end{eqnarray}
and find that $\uv(t) = a \psi_1(t) \uv_1 + b \left[ \psi_1(t) \tilde\uv_2 + \psi_2(t) \uv_1 \right]$, with $\psi_1(t) \uv_1$ being an eigenfunction (the usual normal mode) and ${\bf w}(t) = \psi_1(t) \tilde\uv_2 + \psi_2(t) \uv_1$ is a generalized eigenfunction. This solution 
%${f}_2(t) = \psi_2(t)  $ and $f_1(t) = \left(1+b\left(1+c\right)/a\right)\psi_1(t) - (bc/a)\psi_2(t)$, 
is very similar to what we found in Sec.~\ref{EP_harmonic}, where the generalized eigenfunctions  are composed of $\psi_1$ and $\psi_2$, the two linearly independent solutions.

To conclude, the generalized eigenvectors generate the linearly independent solutions of the differential equations as they should. Notably, without using the generalized eigenvectors one cannot construct a general solution for  the differential equation (in our case, Eq.~\eqref{eq:eom_matrix}).

%Following the same limit procedure used in the analysis fo Eq.~(\ref{eq:EOM}), we that the two eigenvalues at the exceptional point are $-\gamma/2$ and that their associated eigenvectors are
%%%%%%
%\begin{equation}
%\tilde{\wv}_1 = \left( 
%\begin{array}{c}
%1 \\
%-\gamma/2
%\end{array}
%\right) e^{-\gamma t/2},  \qquad \text{and} \qquad 
%\tilde{\wv}_2 = \left( 
%\begin{array}{c}
%t \\
%1 - \frac{1}{2} t
%\end{array}
%\right) e^{-\gamma t/2}
%\end{equation}
%in agreement with refs.{Fernandez} .

%\bibliography{citation_SI.bib}

%apsrev4-2.bst 2019-01-14 (MD) hand-edited version of apsrev4-1.bst
%Control: key (0)
%Control: author (8) initials jnrlst
%Control: editor formatted (1) identically to author
%Control: production of article title (0) allowed
%Control: page (0) single
%Control: year (1) truncated
%Control: production of eprint (0) enabled
%